\documentclass[useAMS]{mn2e}
\usepackage{latexsym,graphicx}

\usepackage{color}

\newcommand\kms{{\rm\,km\,s^{-1}}}
\newcommand\msun{\rm\,M_\odot}
\newcommand\rsun{\rm\,R_\odot}
\newcommand\lsun{\rm\,L_\odot}
\newcommand\hii{H\,{\sc ii} \,}

\newcommand{\MC}{\multicolumn}
\newcommand{\mist}{\texttt{MIST}}
\newcommand{\XMMU}{XMMU\,J051342.6$-$672412}
\newcommand{\MCSNR}{MCSNR\,J0513$-$6724}
\newcommand{\AgeMCSNR}{$\approx4^{+2} _{-1}$\,kyr}
\newcommand{\BSDL}{BSDL\,923}
\newcommand{\TypeBSDL}{B0.7\,III}
\newcommand{\TBSDL}{$T_{\rm eff}=27\pm1$\,kK}
\newcommand{\lgBSDL}{$\log g=3.22\pm0.10$}
\newcommand{\vsinBSDL}{$v\sin i\approx100\pm45\,\kms$}
\newcommand{\eBSDL}{$e=0.158\pm0.061$}
\newcommand{\PBSDL}{$1.280\pm0.006$\,d}
\newcommand{\RBSDL}{$25\pm5\,\rsun$}
\newcommand{\SMABSDL}{$17\pm3\,\rsun$}
\newcommand{\EBV}{$E(B-V)=0.53\pm0.05$~mag}
\newcommand{\Lum}{$\log(L_*/\lsun)=5.46\pm0.10$}
\newcommand{\Alpha}{$\alpha$-elements}
\newcommand{\FAST}{{\sc fastwind}}

\def\apgt{\ {\raise-.5ex\hbox{$\buildrel>\over\sim$}}\ }
\def\aplt{\ {\raise-.5ex\hbox{$\buildrel<\over\sim$}}\ }

\title[Spectroscopy of the LMC HMXB XMMU\,J051342.6$-$672412]
{SALT spectroscopy of the HMXB associated with the LMC supernova remnant MCSNR\,J0513$-$6724}
\author[V. V.~Gvaramadze et al.]
{\fbox{V. V.~Gvaramadze}\thanks{Deceased},$^{1,2,3}$ 
     A. Y.~Kniazev,$^{4,5,1,6}$\thanks{E-mail: a.kniazev@saao.nrf.co.za}
     N.~Castro$^7$ and I. Y.~Katkov$^{8,9,1}$\\
    $^{1}$Sternberg Astronomical Institute, Lomonosov Moscow State University, Universitetskij Pr. 
    13, Moscow 119992, Russia\\
    $^{2}$Space Research Institute, Russian Academy of Sciences, Profsoyuznaya 84/32, Moscow 117997, Russia \\
    $^{3}$E. Kharadze Georgian National Astrophysical Observatory, Abastumani 0301, Georgia \\
    $^{4}$South African Astronomical Observatory, PO Box 9, 7935 Observatory, Cape Town, South Africa \\
    $^{5}$Southern African Large Telescope Foundation, PO Box 9, 7935 Observatory, Cape Town, South Africa \\
    $^{6}$Special Astrophysical Observatory, Nizhnij Arkhyz, Karachai-Circassia, 369167, Russia \\
    $^{7}$Leibniz-Institut f\"ur Astrophysik, An der Sternwarte 16, 14482 Potsdam, Germany \\
    $^{8}$New York University Abu Dhabi, Saadiyat Island, PO Box 129188, Abu Dhabi, UAE \\
    $^{9}$Center for Astro, Particle, and Planetary Physics, NYU Abu Dhabi, PO Box 129188, Abu Dhabi, UAE \\
    }
\begin{document}

\date{Accepted . Received }


\label{firstpage}

\begin{abstract}
\maketitle
We report the results of optical \'echelle spectroscopy with the Southern African Large Telescope (SALT)
of the mass donor star \BSDL\ in the neutron star high-mass X-ray binary \XMMU\ 
associated with the LMC supernova remnant (SNR) \MCSNR. We found that \BSDL\ is a \TypeBSDL\ 
star with double peaked emission lines originating in a circumbinary disk-like structure. 
This classification and the presence of double-peaked emission lines imply that \BSDL\ is a Be star.
Modelling with the stellar atmosphere code \FAST\ was used to derive the effective temperature 
\TBSDL, surface gravity \lgBSDL, projected rotational velocity \vsinBSDL, 
colour excess \EBV, and luminosity \Lum\ of \BSDL, as well as to show that the surface of this star is polluted 
with \Alpha\ (O, Mg and Si) from the supernova ejecta. We found also that the NS is orbiting 
\BSDL\ in an eccentric (\eBSDL) orbit with the orbital period of \PBSDL\ and the 
semi-major axis of \SMABSDL, and the radius of \BSDL\ is \RBSDL. 
We speculate that the NS is embedded in the atmosphere of \BSDL\ either because it was kicked at 
birth towards this star or because of inflation of \BSDL\ caused by the energy input from the supernova 
blast wave. Using long-slit spectroscopy with SALT, we searched for possible signs of the SNR shell in the 
2D spectrum, but did not find them. This lack of detection is consistent with the young age 
(\AgeMCSNR) of \MCSNR, implying that it is still in the adiabatic (non-radiative) phase.
\end{abstract}


\section{Introduction}
\label{sec:int}

MCSNR\,J0513$-$6724 belongs to a rare group of supernova remnants (SNRs) associated with neutron star (NS)
high-mass X-ray binaries (HMXBs). Currently, only six such associations are known. One in our Galaxy: 
SNR\,G322.1+00.0/Cir\,X-1 (Heinz et al. 2013; Linares et al. 2010). Two in the Small Magellanic Cloud:
MCSNR\,J0127$-$7332/SXP\,1062 (H\'enault-Brunet et al. 2012; Haberl et al. 2012; Gvaramadze et al. 2021) 
and MCSNR\,J0103$-$7201/SXP\,1323 (Gvaramadze, Kniazev \& Oskinova 2019a). And the other three in the Large 
Magellanic Cloud (LMC): MCSNR\,J0536$-$6735/CXOU\,J053600.0$-$673507 (Seward et al. 2012; Corbet et al. 
2016; van Soelen et al. 2019), MCSNR\,J0513$-$6724/XMMU\,J051342.6$-$672412 (Maitra et al. 2019) and 
MCSNR\,J0507$-$6847/XMMU\,J050722.1$-$684758 (Maitra et al. 2021). The discovery of these associations is 
a big surprise, since it was believed that the characteristic formation time of HMXBs is several orders of 
magnitude longer than the lifetime of SNRs (e.g. Tauris \& van den Heuvel 2006).
The study of these objects is of great interest for various fields of astrophysics.

Spectroscopic observations of SNRs can be used to establish the type of supernova (SN) explosion that 
led to their formation (e.g. Seitenzahl et al. 2018) or to measure their expansion velocity and thereby 
to determine their age and at what stage of evolution they are (e.g. Lozinskaya 1992; Gvaramadze et al. 
2019a, 2021). The latter in turn provides constraints on the models proposed to explain the origin of 
their associated HMXBs. In particular, knowledge of the age of HMXBs is necessary for constructing 
models of magnetic field and spin evolution of young NSs in binary systems (e.g. Shakura et al. 2012; 
Fu \& Li 2012; Christodoulou, Laycock \& Kazanas 2018; Ho et al. 2020; Wang \& Tong 2020).

Spectroscopic observations of mass donor stars in HMXBs, in turn, are necessary to determine whether they 
are Be or supergiant HMXBs (e.g. van Kerkwijk, van Oijen \& van den Heuvel 1989; McBride et al. 2008), 
to measure chemical abundances on their surfaces, and potentially to detect abundance anomalies caused by 
the pollution by SN ejecta (cf. Israelian et al. 1999; Gvaramadze et al. 2017). Spectroscopic observations 
are also needed to determine orbital parameters of HMXBs, such as their orbital periods, eccentricities, 
mass functions, and so on (e.g. van Kerkwijk et al. 1985; Townsend et al. 2011). The latter parameters can 
be used to infer SN kick velocities imparted to new-born NSs, to determine orbital parameters of pre-SN 
binaries and evolution of their post-SN orbits (e.g. Hills 1983; Tauris et al. 1999), and ultimately to 
find out final fates of HMXBs (e.g. Bhattacharya \& van den Heuvel 1991; van den Heuvel 2019).

In this paper, we report the results of optical spectroscopic observations of the mass donor star 
BSDL\,923 in the HMXB XMMU\,J051342.6$-$672412 associated with the LMC SNR MCSNR\,J0513$-$6724 
with the Southern African Large Telescope (SALT). In Section\,\ref{sec:obs}, we review what is known 
about XMMU\,J051342.6$-$672412 and its host SNR. The SALT observations are described in 
Section\,\ref{sec:spe} and analysed in Section\,\ref{sec:res}. In Section\,\ref{sec:tess}, we analyse
the TESS (Transiting Exoplanet Survey Satellite) data for XMMU\,J051342.6$-$672412. The obtained results 
are discussed in Section\,\ref{sec:dis} and summarized in Section\,\ref{sec:sum}.

\begin{table}
  \centering{\caption{Properties of BSDL\,923.}
  \label{tab:det}
 \begin{tabular}{lcc}  	
 \hline
  Spectral type & B0.7\,III & This paper \\
  RA(J2000) & $05^{\rm h} 13^{\rm m} 42\fs60$ & 2MASS \\ 	
  Dec.(J2000) & $-67\degr 24\arcmin 10\farcs1$ & 2MASS \\ 
  $U$ (mag) & $12.97\pm0.03$ & Zaritsky et al. (2004) \\
  $B$ (mag) & $13.75\pm0.03$ & Zaritsky et al. (2004) \\
  $V$ (mag) & $13.45\pm0.04$ & Zaritsky et al. (2004) \\
  $I_{\rm c}$ (mag) & $13.11\pm0.04$ & Zaritsky et al. (2004) \\
  $J$ (mag) & $12.93\pm0.03$ & 2MASS \\  
  $H$ (mag) & $12.84\pm0.04$ & 2MASS \\
  $K_{\rm s}$ (mag) & $12.69\pm0.04$ & 2MASS \\
  3.6 \micron \, (mag) & $13.01\pm0.03$ & IPAC 2009 \\ 
  4.5 \micron \, (mag) & $12.93\pm0.03$ & IPAC 2009 \\
  5.8 \micron \, (mag) & $12.82\pm0.07$ & IPAC 2009 \\
  8 \micron \, (mag)   & $12.36\pm0.14$ & IPAC 2009 \\ 
  \hline
 \end{tabular}
}
\end{table}

\begin{table*}
\caption{Journal of the SALT observations.}
\label{tab:log}
\begin{tabular}{llccccc} \hline
Date & Grating & Exposure & Spectral scale      & Slit/Fiber & Seeing   & Spectral range \\
     &         & (sec)    & (\AA\,pixel$^{-1}$) & (arcsec)   & (arcsec) & (\AA) \\
     \hline
     \MC{7}{c}{Long-slit observations} \\
     \hline
2019 October 10  & PG2300 & 1400$\times$2 &  0.26  & 1.50   & 2.2 & 6030$-$6870  \\
     \hline
     \MC{7}{c}{\'Echelle observations} \\
     \hline
2019 October 8   & LR     & 2800$\times$1 &  0.042 & 2.23   & 1.1 & 3840$-$8800 \\
2019 October 11  & LR     & 3000$\times$1 &  0.042 & 2.23   & 1.2 & 3840$-$8800 \\
2019 October 14  & LR     & 3000$\times$1 &  0.042 & 2.23   & 2.0 & 3840$-$8800 \\
2020 October 06  & MR     & 3000$\times$1 &  0.042 & 2.23   & 1.1 & 3840$-$8800 \\
2020 October 11  & MR     & 3000$\times$1 &  0.042 & 2.23   & 2.5 & 3840$-$8800 \\
2020 October 18  & MR     & 3000$\times$1 &  0.042 & 2.23   & 1.5 & 3840$-$8800 \\
2020 October 23  & MR     & 3000$\times$1 &  0.042 & 2.23   & 1.6 & 3840$-$8800 \\
2020 October 31  & MR     & 3000$\times$1 &  0.042 & 2.23   & 2.0 & 3840$-$8800 \\
2020 November 11 & MR     & 3000$\times$1 &  0.042 & 2.23   & 3.1 & 3840$-$8800 \\
2020 November 27 & MR     & 3000$\times$1 &  0.042 & 2.23   & 2.8 & 3840$-$8800 \\
2020 December 10 & MR     & 3000$\times$1 &  0.042 & 2.23   & 2.5 & 3840$-$8800 \\ \hline
\end{tabular}
\end{table*}

\section{LMC SNR MCSNR\,J0513$-$6724 and its associated HMXB XMMU\,J051342.6$-$672412}
\label{sec:obs}

Maitra et al. (2019) detected an X-ray source of luminosity of $L_{\rm X}\approx 7\times10^{33} 
\, {\rm erg} \, {\rm s}^{-1}$ at the geometrical centre of the candidate 
SNR MCSNR\,J0513$-$6724 (Bozzetto et al. 2017) using data from {\it XMM-Newton}. An inspection 
of the X-ray light curve of this source revealed 4.4\,s tentative pulsations, which were attributed to 
the NS spin period. Maitra et al. (2019) also identified an optical counterpart to the X-ray 
source with the star BSDL\,923\footnote{Indicated in the SIMBAD data base as a post-AGB star 
candidate.} ($V\approx13$\,mag), whose spectral (B2.5) type and luminosity class (Ib) were 
obtained by Dachs (1972) from the $U-B$ and $B-V$ colours of the star and its membership of the 
LMC. The supergiant luminosity class of BSDL\,923 was also derived by Maitra et al. (2019) on 
the basis of photometric measurements in a much wider spectral range. This suggests that the 
detected X-ray source is a supergiant HMXB. It was named XMMU\,J051342.6$-$672412 by Ho et al. 
(2020) and we will use this name hereafter.

The Optical Gravitational Lensing Experiment (OGLE) $I$-band light curve of BSDL\,923 displays a 
nearly sinusoidal variability by about $\pm0.15$\,mag over $\approx19$\,yr and $\approx0.05$\,mag 
variations within the half-yearly visibility windows (Maitra et al. 2019). After removing the 
long-term trend from the light curve and using the Lomb-Scargle (LS; Lomb 1976; Scargle 1982) 
technique, Maitra et al. (2019) found two strong peaks in the LS periodogram at 1.8025 and 
2.2324\,d. The latter (strongest) peak was interpreted as the likely orbital period of the HMXB.

In X-rays the SNR appears as an incomplete shell whose outer contour could be fitted by 
an ellipse with the semi- major and semi-minor axes of $\approx60$ and 50 arcsec respectively (see 
fig.\,1 in Maitra et al. 2019). At the distance to the LMC of 49.9 kpc (Pietrzy\'nski et al. 2013), 
the linear size of the SNR is $\approx14 \, {\rm pc} \times 12$ pc. From the X-ray spectral analysis 
of MCSNR\,J0513$-$6724, Maitra et al. (2019) determined the plasma temperature of $2.2^{+1.2} _{-1.1}$ 
keV. Assuming that the SNR is the Sedov phase and using the radius of the SNR of $\approx14$ pc, one 
can estimate the SNR expansion velocity and age of, respectively, $\approx1400 ^{+300} _{-400} \, \kms$ 
and $\approx4000 ^{+2000} _{-1000}$\,yr (cf. Maitra et al. 2019). This makes XMMU J051342.6$-$672412 
the youngest NS HMXB associated with a SNR known to date.

Maitra et al. (2019) also reported detection of MCSNR\,J0513$-$6724 in the Australian Square 
Kilometre Array Pathfinder (ASKAP) survey of the LMC at 888\,MHz (Pennock et al.\,2021). 
Their fig.\,2 shows that the 
brightest radio emission spatially coincides with the region of the brightest and hardest X-ray emission,
which supports the SNR interpretation of MCSNR\,J0513$-$6724. The further support for this interpretation
comes from the spectral index of the radio emission of $\alpha=-0.68\pm0.04$ (Maitra et al. 2019).

Table\,\ref{tab:det} summarizes some properties of BSDL\,923. The coordinates and the $JHK_s$ photometry 
are from the 2MASS All-Sky Catalog of Point Sources (Cutri et al. 2003). The $UBVI$ photometry is from 
the Magellanic Clouds Photometric Survey (Zaritsky et al. 2004). The 3.6, 4.5, 5.8 and 8 \micron \, 
photometry is from the SAGE LMC and SMC IRAC Source Catalog (IPAC 2009) (Meixner et al. 2006).

\section{SALT spectroscopy}
\label{sec:spe}

To study BSDL\,923 and to characterize the binary orbit, we obtained 11 spectra with the 
High-Resolution Spectrograph (HRS; Barnes et al. 2008; Bramall et al. 2010, 2012; Crause et al. 
2014) on SALT (Buckley, Swart \& Meiring 2006; O'Donoghue et al. 2006) in 2019--2020. Using this 
dual beam, fibre-fed \'echelle spectrograph, we obtained three 
spectra in the low resolution (LR) mode (the resolving power of $R=14\,000-15\,000$) in 2019 
and other 8 spectra in the medium resolution (MR) mode ($R=36\,500-39\,000$; Kniazev et al. 
2019) in 2020. In both modes the object and sky fibres have a 2.23 arcsec diameter. The resulting 
spectra in the blue and red arms cover the spectral range of $\approx3700-8800$\,\AA. All 
but the first spectrum were obtained with an exposure of 3000\,s, and the exposure of the first 
spectrum was 2800\,s. The seeing during these observations was in the range from 1.1 to 3.1 arcsec.
Spectra of a ThAr lamp and spectral flats were obtained for both modes during a weekly set of HRS 
calibrations.

Due to lower resolution and wider entrance slit, the LR HRS spectra have a better signal-to-noise 
(S/N) ratio (S/N$\sim90$ at the maximum of an \'echelle order in the spectral region around the 
H\,$\beta$ line) than the MR HRS ones (S/N$\sim60$ for the same spectral region). Correspondingly, 
the MR spectra were used only to measure the radial velocity of the star and to search for changes in
profiles of strong spectral lines, while the LR spectra were also used for spectral classification
and modelling.

Also, we carried out long-slit spectroscopic observations of the field containing MCSNR\,J0513$-$6724
using the SALT Robert Stobie Spectrograph (RSS; Burgh et al. 2003; Kobulnicky et al. 2003). The 
observations were intended to search 
for possible signs of the SNR shell in the 2D spectrum and/or interaction of the SN blast wave with
the local interstellar medium. The RSS spectra were taken on 2019 October 10 with a 1.5 arcsec slit 
(placed on BSDL\,923 at position angle of PA=20\degr) and the PG2300 grating, covering the spectral 
range of 6030--6870\,\AA. This setup provides a spectral resolution of FWHM of $2.20\pm0.25$~\AA. Two 
1400\,s exposures were obtained under seeing conditions of 2.2 arcsec. For wavelength calibration of 
the spectra an Xe lamp arc spectrum was taken immediately after the science frames. A spectrophotometric 
standard star was observed with the same spectral setup for the relative flux calibration.
An HRS spectrophotometric standard is observed once per month as a part of the HRS Calibration Plan.
All HRS spectrophotometric standards have spectral distributions known in steps of 3--4~\AA.
We need to note here that SALT is a telescope with a variable pupil, 
so that the illuminating beam changes continuously during 
the observations. This makes absolute flux calibration impossible even when using 
spectrophotometric standards.
However, the relative energy distributions are very accurate, 
especially as the SALT has an atmospheric dispersion compensator (ADC).

The obtained spectra were first reduced using the SALT science pipeline (Crawford et al. 2010). The 
further reduction of the \'echelle spectra was performed using the HRS pipeline described in Kniazev, 
Gvaramadze \& Berdnikov (2016) and Kniazev et al. (2019). The long-slit spectra were further reduced as 
described in Kniazev (2022). 

The journal of the SALT observations is given in Table\,\ref{tab:log}.

\begin{figure*}
    \begin{center}
        \includegraphics[width=12cm,angle=0]{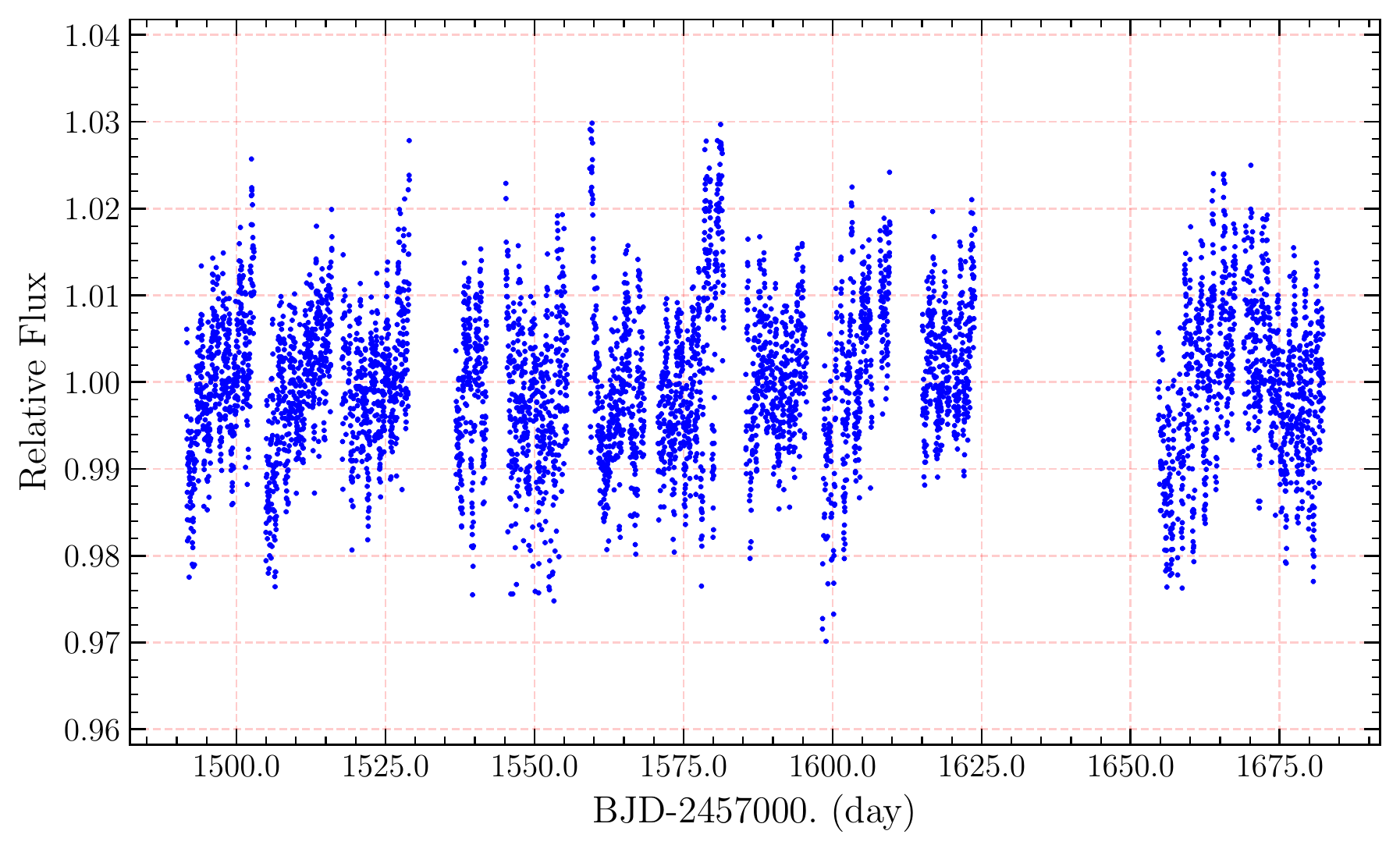}
        \includegraphics[width=12cm,angle=0]{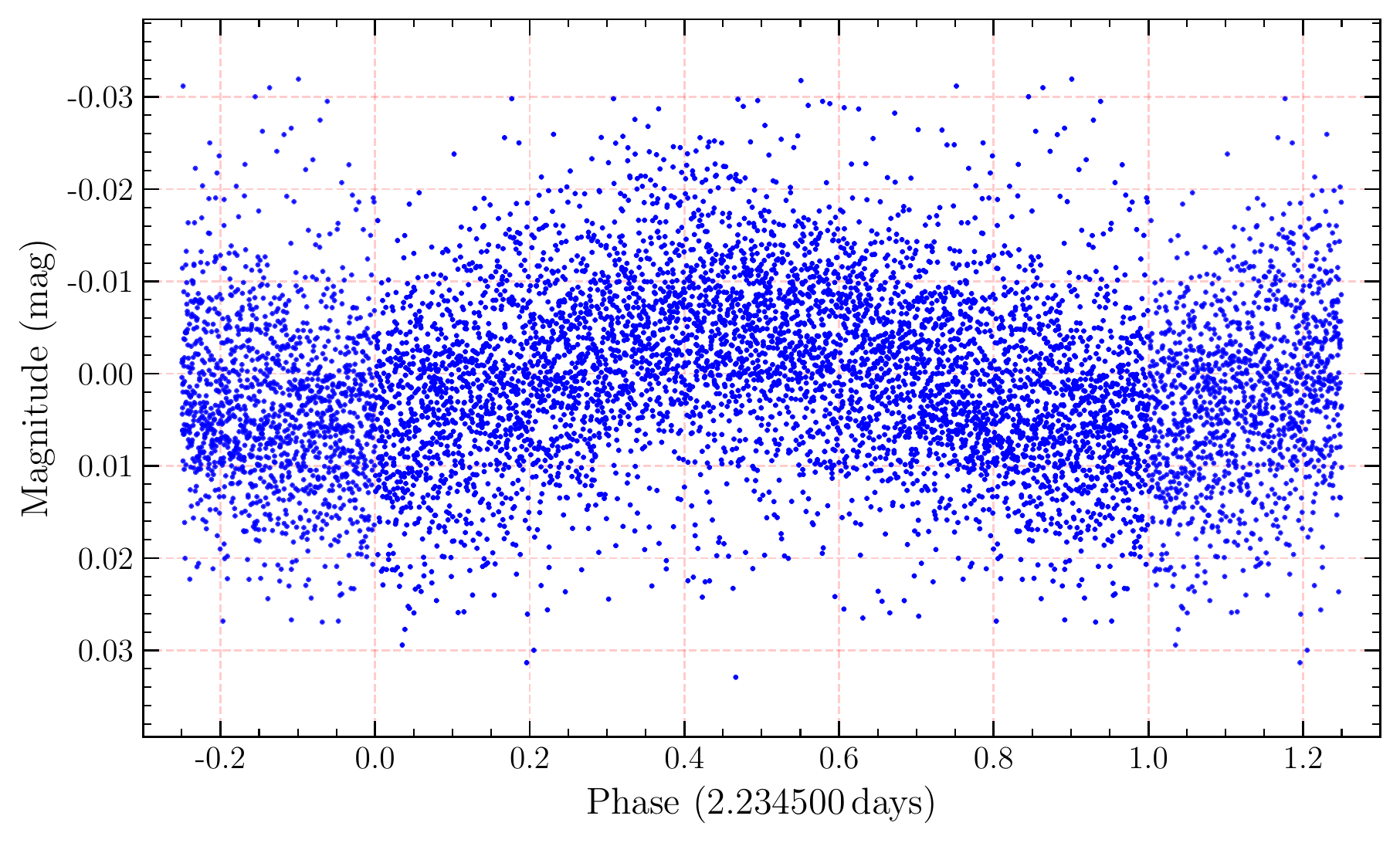}
    \end{center}
    \caption{Top panel: The TESS light curve of BSDL\,923 from the QLP pipeline (Huang et al. 2020ab).
    Bottom panel: The TESS light curve folded with a period of 2.2324~d.} 
    \label{fig:TESS_data}
\end{figure*}

\begin{figure}
\begin{center}
\includegraphics[width=8.5cm,angle=0]{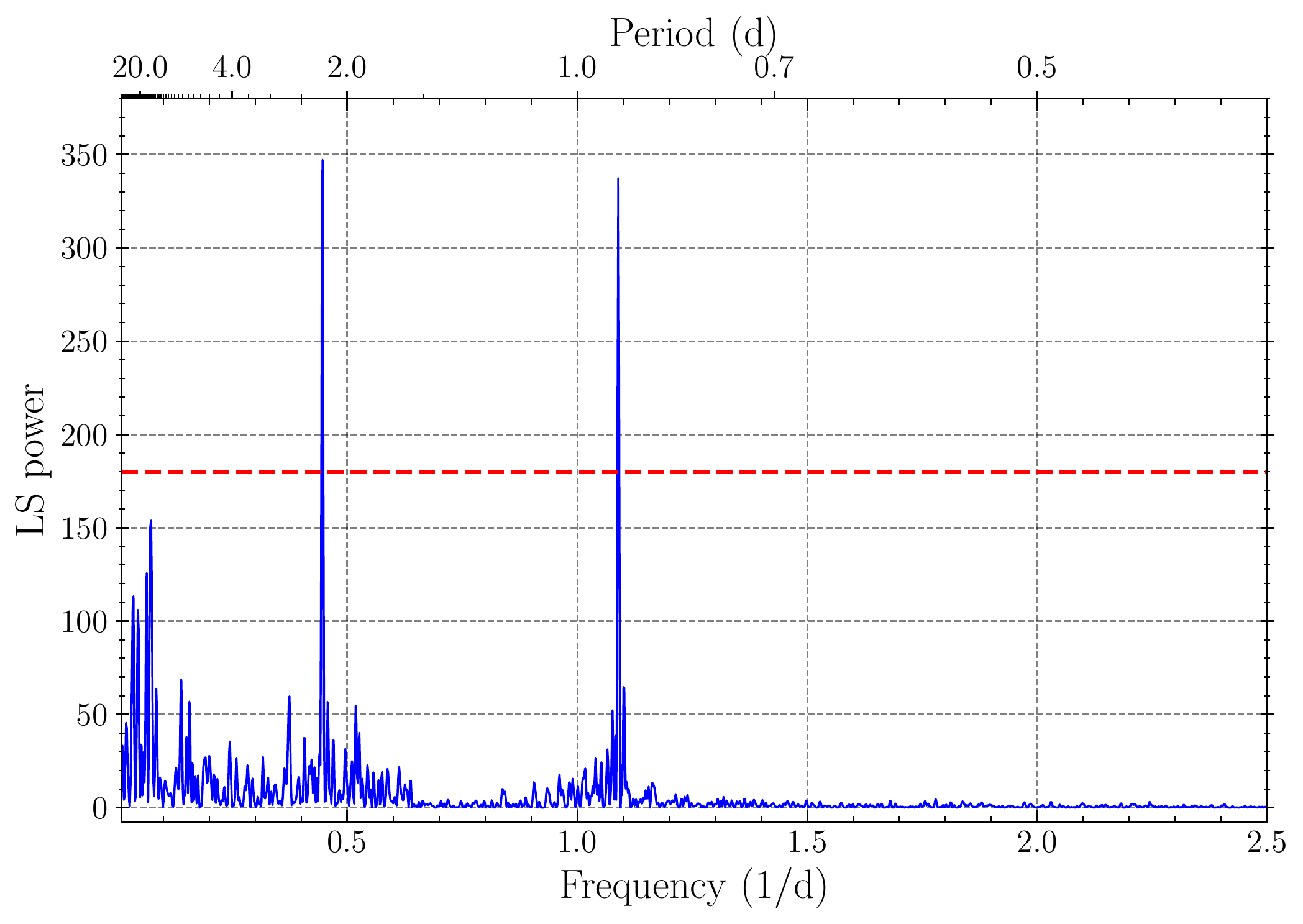}
\end{center}
\caption{The LS periodogram of the TESS light curve of BSDL\,923. The two strongest 
spikes correspond to periods of $2.2345\pm0.0002$ and $0.9165\pm0.0001$~d. The dashed (red) line 
shows the $5\sigma$ confidence level.
} 
\label{fig:LS_diag}
\end{figure}

\section{TESS}
\label{sec:tess}

TESS (Ricker et al. 2014) is a NASA Explorer-class mission 
aiming to image nearly the entire sky to search for exoplanets using the transit method. The standard 
TESS cadence is 30\,min throughout a 30\,d fixed-pointing ``sector''. The large part of reduced TESS 
data is publicly released and can be obtained from the Mikulski Archive for Space Telescopes 
(MAST)\footnote{https://mast.stsci.edu}. In the current work, we use TESS light curves from the MIT 
quick-look pipeline (``QLP''; Huang et al. 2020ab)\footnote{See also https://archive.stsci.edu/hlsp/qlp 
for more details.}. BSDL\,923 was observed in the TESS sectors 7, 8, 9, 10, 11, and 13, which covered 
about half-year period with some holes in data related to data quality.

All available TESS data with quality flag 0 are shown in the top panel of Fig.~\ref{fig:TESS_data}. 
We analysed these photometric data in the same way as it was done in Maryeva et al.\,(2020). 
The Lomb-Scargle (LS) periodogram (Lomb 1976; Scargle 1982) using the {\sc midas} context 
{\sc tsa} of these data (see Fig.~\ref{fig:LS_diag}) shows two significant 
peaks at periods of 2.2345$\pm$0.0002 and 0.9165$\pm$0.0001\,d. 
The first of these periods is almost equal to the period of 2.2324\,d obtained by 
Maitra et al.\,(2019) from the analysis of the OGLE data (see Section\,\ref{sec:obs}), 
while their second period of 1.8025\,d is just the second harmonic 
of the TESS peak at 0.9165\,d; note that the one-day sampling period of the OGLE data did 
not allow Maitra et al.(2019) to detect periods shorter than 1\,d. 
In Section~\ref{sec:orb}, we show that none of the periods detected in the OGLE and 
TESS data are associated with the orbital period of XMMU\,J051342.6$-$672412.
The bottom panel of Fig.~\ref{fig:TESS_data} shows the TESS light curve folded with a period of 2.2324~d.
\begin{figure*}
\begin{center}
\includegraphics[width=15.5cm,angle=0]{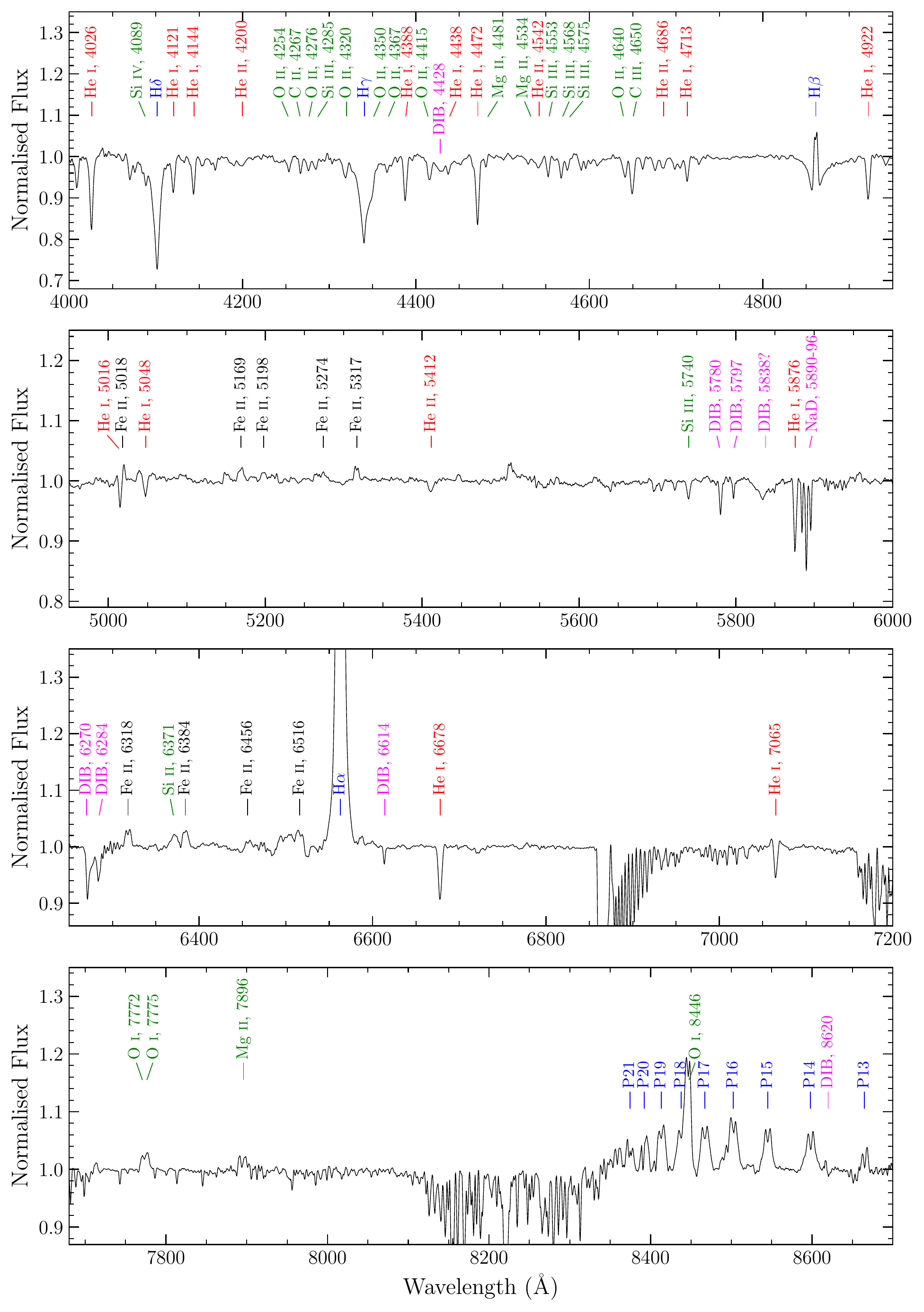}
\end{center}
\caption{Portions of the normalized merged LR HRS spectra of BSDL\,923 degraded to $R=4000$ with main 
lines and diffuse interstellar bands (DIBs) indicated. The feature at 5520\,\AA \, is due to merging of 
\'echelle spectra obtained with the blue and red arms of HRS.
} 
\label{fig:bsdl}
\end{figure*}

\begin{figure*}
\begin{center}
\includegraphics[width=5.5cm,angle=0]{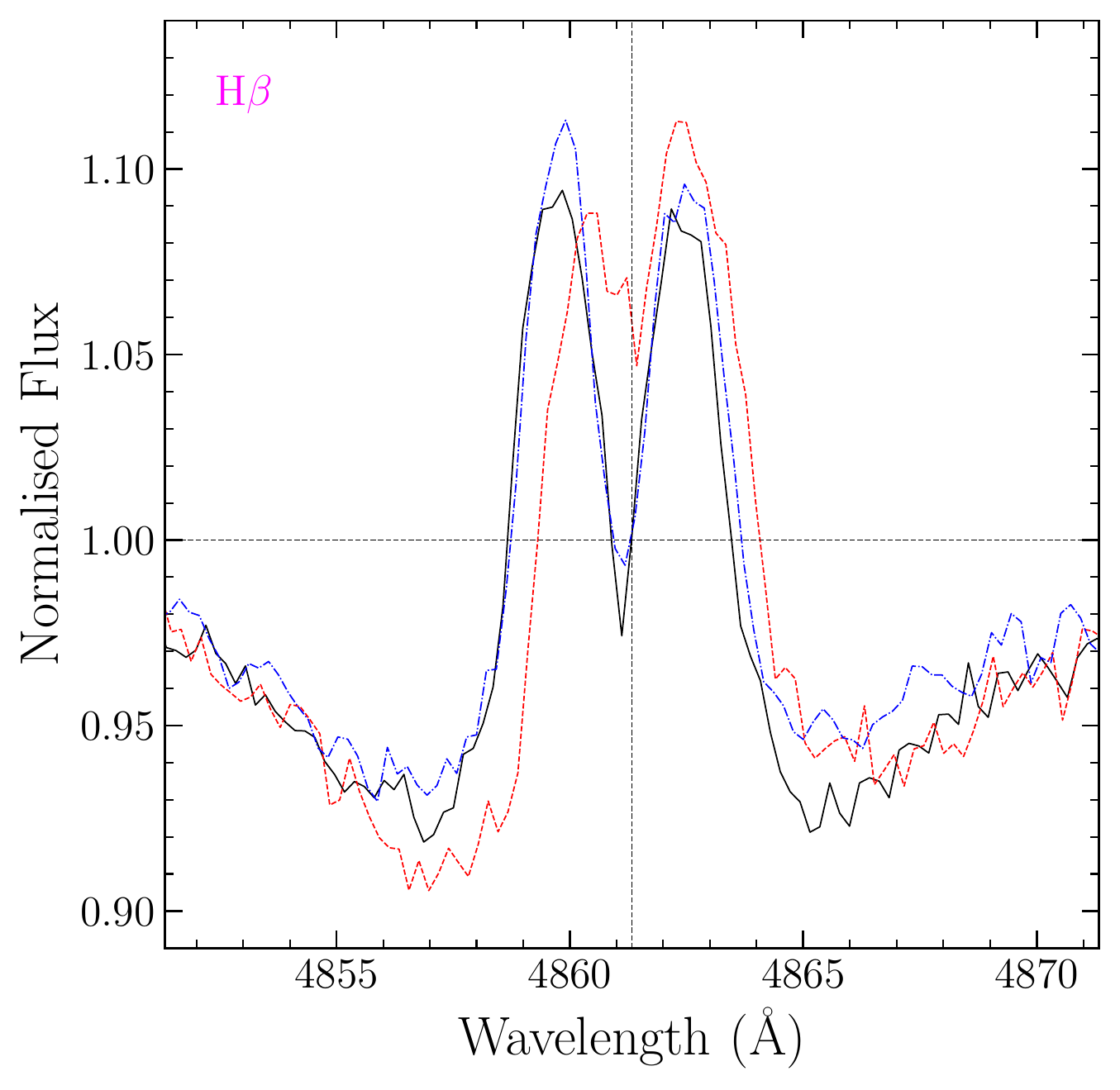}
\includegraphics[width=5.5cm,angle=0]{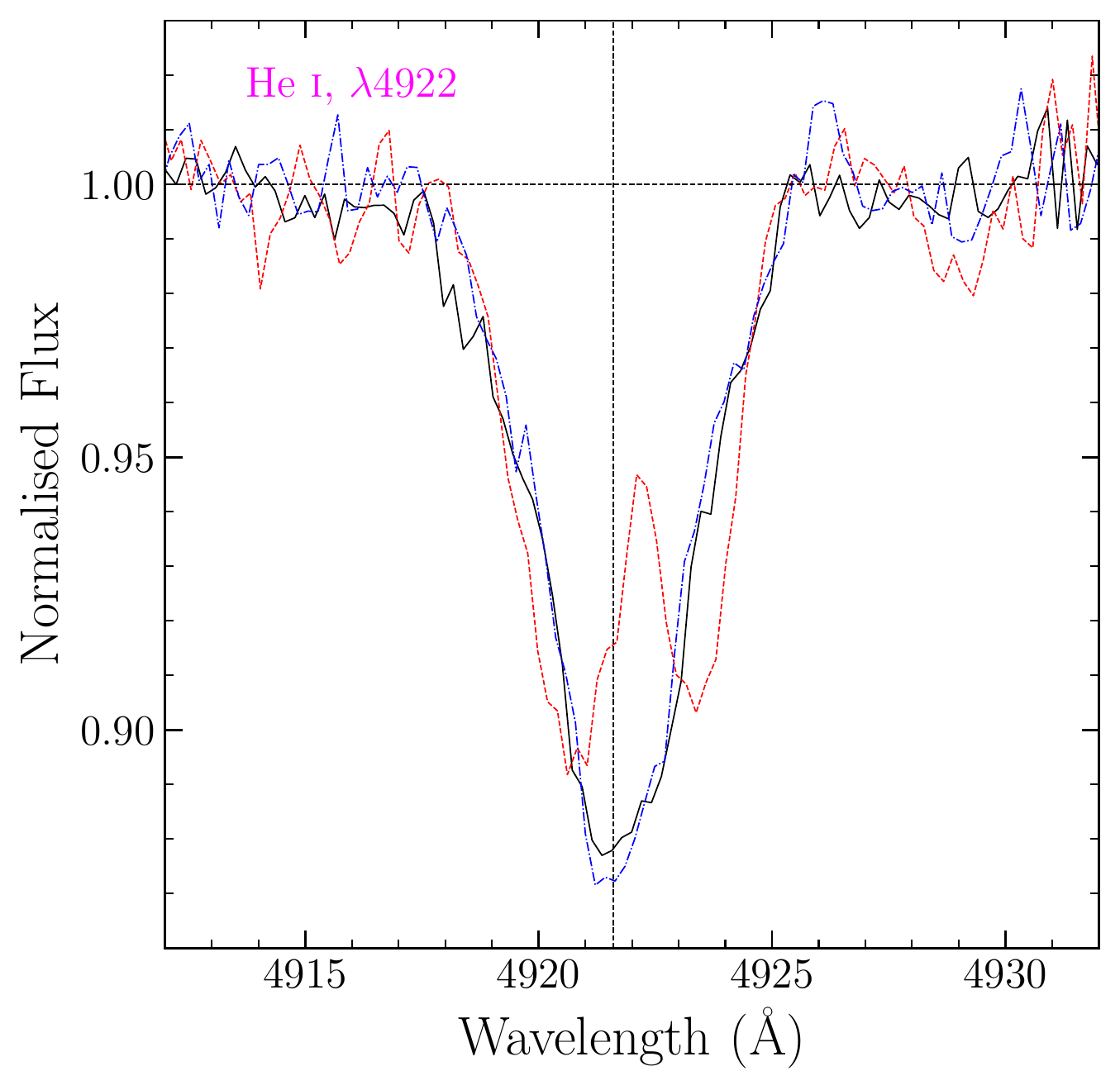}
\includegraphics[width=5.5cm,angle=0]{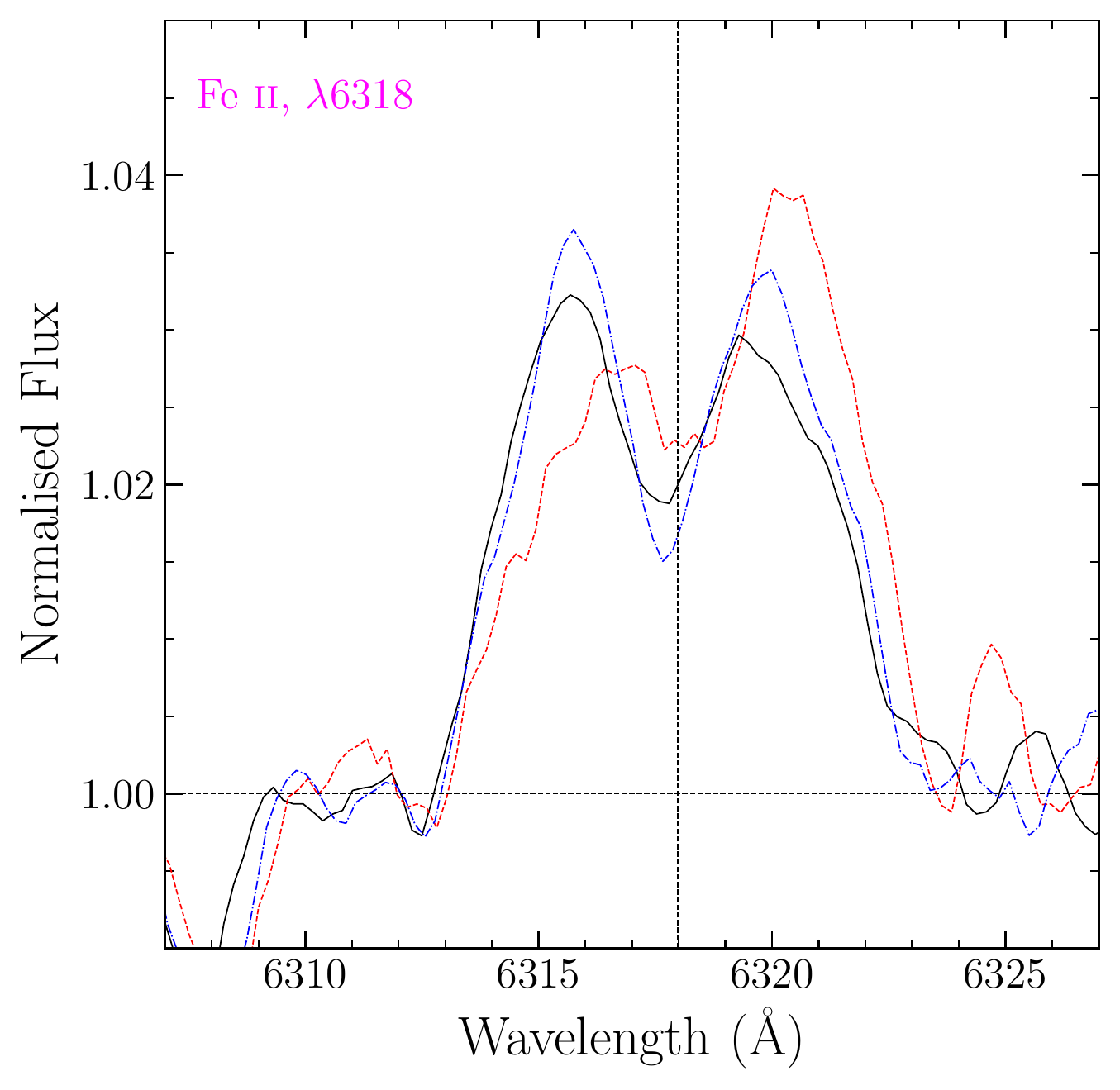}
\end{center}
\caption{Profile variability of three representative lines in the LR HRS spectra of BSDL\,923 obtained 
on 2019 October 8 (black solid line), 2019 October 11 (red dashed line) and 2019 October 14 (blue 
dot-dashed line). The lines are shown in the reference frame of BSDL\,923.} 
\label{fig:lines}
\end{figure*}

\section{Results of spectroscopy}
\label{sec:res}

\subsection{Spectral classification of BSDL\,923}
\label{sec:cla}

The spectrum of BSDL\,923 (see Fig.\,\ref{fig:bsdl}) is dominated by the Balmer and He\,{\sc i} lines, 
of which all He\,{\sc i} and the higher-order Balmer lines are in absorption, while the H\,$\alpha$ and 
H\,$\beta$ lines are in emission. The latter line has a double-peaked profile and is surrounded on both 
sides by photospheric absorption line wings. The spectrum also shows weak absorption lines of He\,{\sc ii} 
at 4200, 4542, 4686 and 5412\,\AA. Other double-peaked emission lines in the spectrum are due to 
Fe\,{\sc ii}. The near-infrared part of the spectrum shows lines of the Paschen series and persistent 
lines of neutral oxygen (O\,{\sc i} $\lambda\lambda$7772-5, 8446), all of which are in emission and 
double-peaked as well. The double-peaked shape of the lines indicates that they originate in a flattened 
rotating structure, like an excretion disk around a Be star. 

Comparison of the three LR HRS spectra revealed obvious changes in both emission and absorption 
line profiles. These changes are illustrated in Fig.\,\ref{fig:lines} by the examples of three 
representative lines (H\,$\beta$, He\,{\sc i} $\lambda$4922 and Fe\,{\sc ii} $\lambda$6318), which are 
shown in the reference frame of BSDL\,923. We found that profiles of absorption lines in the first and 
third spectra do not differ much from each other, while in the second spectrum a red-shifted emission 
component has appeared in their cores. 

To classify BSDL\,923, we visually compared its spectrum with spectra of B stars in the LMC from 
Evans et al. (2015). For this, we merged the first and third LR HRS spectra after their correction 
for heliocentric motion, radial velocity variations (see Section\,\ref{sec:orb}) and redshift of the 
LMC, and then smoothed and degraded the resulting spectrum to the resolving power of $R=4000$ used 
in Evans et al. (2015). We used only these two spectra because they are less affected by
emission from the circumstellar material. Parts of the resulting spectrum, after normalization, 
are shown in Fig.\,\ref{fig:bsdl}.

The presence of He\,{\sc ii} lines indicates the spectral type earlier than B1 star, while the 
non-detection of the Si\,{\sc iv} $\lambda4116$ line and the nearly equal strength of the Si\,{\sc 
iv} $\lambda$4089 and Si\,{\sc iii} $\lambda$4553 lines implies that BSDL\,923 is either a B0.7 giant 
or a B0.5-B0.7 dwarf star (Evans et al. 2015; see also their figs A1--A3). The giant luminosity class 
is more preferable in view of the surface gravity derived for BSDL\,923 in Section\,\ref{sec:par}. We
therefore conclude that BSDL\,923 is a B0.7\,III star. Although this classification and the presence 
of double-peaked emission lines imply that BSDL\,923 is a Be star and that XMMU\,J051342.6$-$672412 is 
a Be HMXB, the reality appears to be more interesting (see Section\,\ref{sec:dis}). 

\begin{table*}
    \centering\caption{Heliocentric radial velocity, $V_{\rm hel}$, of BSDL\,923 and some parameters 
    of the H\,$\alpha$ line measured at different orbital phases. $\Delta V$ is the difference between 
    the observed values of $V_{\rm hel}$ and the model fit of the radial velocity curve of BSDL\,923 to 
    the Keplerian orbit.}
    \label{tab:rad}
    \begin{tabular}{lccccccc} 
\hline
Date             & JD$-$2450000 & $V_{\rm hel}$ & $\Delta V$ & Phase & EW(H\,$\alpha$) 
& FWHM(H\,$\alpha$) & $V_{\rm hel}$(H\,$\alpha$) \\
                 &     (d)     & $(\kms)$      & $(\kms)$ &         & (\AA) & (\AA) & $(\kms)$  \\
\hline                            
2019 October 8   & 8765.52109  & $313.8\pm0.5$ & 0.0022   & 0.76206 & $-17.5\pm0.4$ & $5.60\pm0.02$ & $312.47\pm0.21$ \\
2019 October 11  & 8768.51130  & $287.8\pm0.6$ &$-2.4543$ & 0.09877 & $-18.3\pm0.4$ & $5.53\pm0.01$ & $306.53\pm0.11$ \\
2019 October 14  & 8771.52773  & $312.2\pm0.4$ & 0.2653   & 0.45597 & $-18.0\pm0.5$ & $5.10\pm0.01$ & $310.36\pm0.12$ \\
2020 October 06  & 9129.55857  & $298.9\pm0.6$ & 0.8967   & 0.24050 & $-16.5\pm0.3$ & $5.65\pm0.01$ & $306.18\pm0.17$ \\
2020 October 11  & 9134.52554  & $293.7\pm0.6$ & 2.8017   & 0.12196 & $-15.7\pm0.2$ & $6.02\pm0.02$ & $306.20\pm0.26$ \\
2020 October 18  & 9141.56463  & $316.5\pm0.5$ &$-0.0628$ & 0.62268 & $-16.2\pm0.2$ & $5.52\pm0.01$ & $309.26\pm0.15$ \\
2020 October 23  & 9146.45694  & $311.3\pm0.3$ &$-0.1078$ & 0.44580 & $-15.6\pm0.2$ & $5.47\pm0.02$ & $308.26\pm0.24$ \\
2020 October 31  & 9154.49222  & $316.5\pm0.7$ & 1.1884   & 0.72499 & $-14.9\pm0.2$ & $6.86\pm0.07$ & $307.43\pm0.90$ \\
2020 November 11 & 9165.43836  & $298.3\pm0.8$ &$-2.5654$ & 0.27890 & $-16.2\pm0.2$ & $5.58\pm0.02$ & $306.97\pm0.21$ \\
2020 November 27 & 9181.42116  & $311.5\pm0.8$ &$-1.9786$ & 0.76872 & $-15.3\pm0.2$ & $6.16\pm0.05$ & $307.19\pm0.52$ \\
2020 December 10 & 9194.40253  & $303.2\pm1.2$ & 1.9204   & 0.91306 & $-17.5\pm0.2$ & $5.47\pm0.02$ & $300.16\pm0.23$ \\
\hline
    \end{tabular}
\end{table*}

\begin{figure*}
\begin{center}
\includegraphics[width=12cm,angle=0]{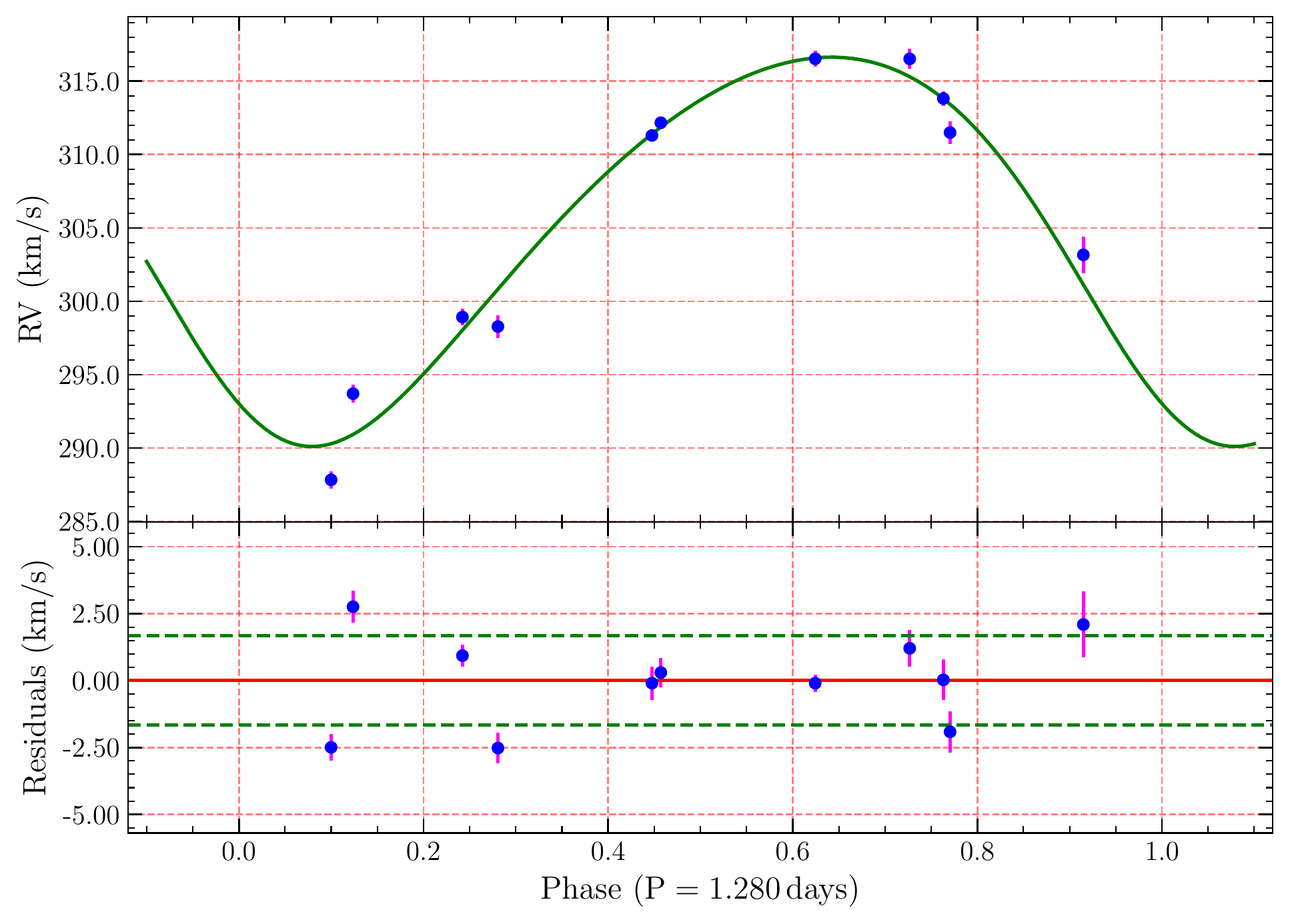}
\end{center}
\caption{Upper panel: Radial velocity curve of BSDL\,923 overlayed with a sine fit with a period of 
$1.280\pm0.006$\,d to the heliocentric radial velocities listed in Table\,\ref{tab:rad}. Bottom panel: 
residuals of the fit, with a rms of $1.66 \, \kms$.} 
\label{fig:orb}
\end{figure*}
%
\begin{table*}
    \centering
    \caption{Orbital parameters of BSDL\,923. 
    }
    \begin{tabular}{lc}
        \hline
        Parameter                                              &  Value     \\ 
        \hline
        The closest epoch at periastron $T_{\rm p}$ (d)        &  2458764.5459$\pm$0.1026  \\
        The superior conjunction phase $\phi _{\rm s}$         &  0.318                    \\
        The inferior conjunction phase $\phi _{\rm i}$         &  0.893                    \\
        Orbital period $P$ (d)                                 &  1.280$\pm$0.006          \\
        Eccentricity $e$                                       &   0.158$\pm$0.061         \\
        Radial velocity semi-amplitude of the primary $K$ ($\kms$) & 13.27$\pm$0.84        \\
        Mass function $f(m)=(1-e^2)^{3/2}P K^3/2\upi G$ ($\msun$) & $(2.98^{+0.71} _{-0.63})\times10^{-4}$ \\
        The longitude of the periastron $\omega$  (degrees)    &   141.72$\pm$22.76        \\
        Systemic heliocentric velocity $\gamma$ ($\kms$)       &   305.02$\pm$0.74         \\
        Root-mean-square residuals of Keplerian fit ($\kms$)   &       1.66                \\
        \hline
    \end{tabular}
    \label{tab:orb}
\end{table*}

\subsection{Radial velocity measurements and the orbital solution to XMMU\,J051342.6$-$672412}
\label{sec:orb}

To characterize the binary orbit, we measured the heliocentric radial velocity, $V_{\rm hel}$, of 
BSDL\,923 using all 11 HRS spectra. For this, we utilized the {\sc fbs} (Fitting Binary Stars) software 
developed by our group and described in Kniazev et al. (2020) and Kniazev (2020).
The obtained values of $V_{\rm hel}$ are given in Table\,\ref{tab:rad} along with dates of observations. 
To this table we also added EWs, FWHMs and heliocentric radial velocities of the H\,$\alpha$ line 
measured in each HRS spectrum. 

Table\,\ref{tab:rad} shows that $V_{\rm hel}$ is variable on a time scale of several days. To check 
whether this variability is described by one of the three possible orbital periods, 0.9165, 
1.8025 or 2.2345\,d, derived from the analysis of the OGLE and TESS data (see Sections\,\ref{sec:obs} 
and \ref{sec:tess}), we fit the available data points to a binary model assuming that the binary components 
are point masses moving in Keplerian orbits. For that {\sc fbs} software was used as well.
The results of the fit are shown in Fig.\,\ref{fig:orb} and summarized in Table\,\ref{tab:orb}. 

From Table\,\ref{tab:orb} it follows that none of the three periods fit the radial velocity 
measurements. It means that optical variations are not due to the system binarity,
but have another reason, for example, due to the rotational modulation of B-star (e.g. Balona 2016)
or due to the rotation of the circumbinary material (see Section\,\ref{sec:dis} for more details).

Table\,\ref{tab:orb} also shows that the systemic velocity of the binary system, $\gamma$, 
is equal to the systemic velocity of the local interstellar medium detected in the RSS spectrum (see 
Section\,\ref{sec:snr}) and agrees well with the local systemic velocity ($\approx300 \, \kms$) of the 
LMC (Kim et al. 1998), implying that the binary received a low or zero birth kick along our line of sight. 

\begin{figure}
\begin{center}
\includegraphics[width=0.48\textwidth,angle=0]{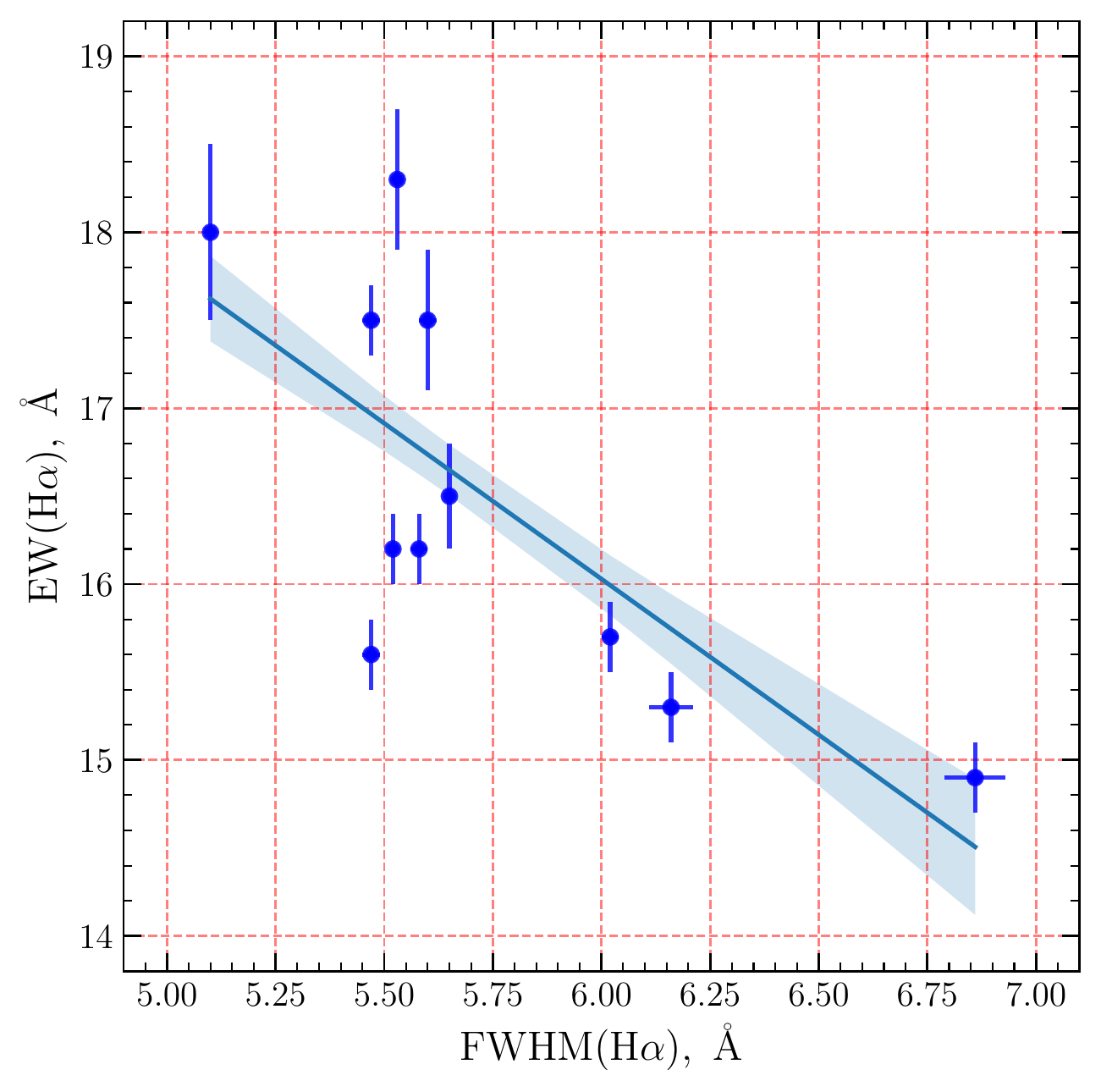}
\end{center}
\caption{EW of the H\,$\alpha$ line as a function of its FWHM. The solid (blue) line is the 
least-squares linear fit. The shaded area is the $1\sigma$ uncertainty of the fit.} 
\label{fig:EW}
\end{figure}

Moreover, Table\,\ref{tab:rad} shows that despite the orbital motion of BSDL\,923, the heliocentric radial 
velocity of the H\,$\alpha$ emission line (originating in the circumstellar medium) remains approximately 
equal to the systemic velocity of the binary system (we discuss this remarkable fact in the 
Section\,\ref{sec:dis}). In addition, we did not find any correlation between EW or FWHM of the H\,$\alpha$ 
line and the orbital phase, but found that EW tends to decrease with increasing FWHM (see Fig.\,\ref{fig:EW}).

\begin{figure*}
\begin{center}
\includegraphics[width=16cm,angle=0]{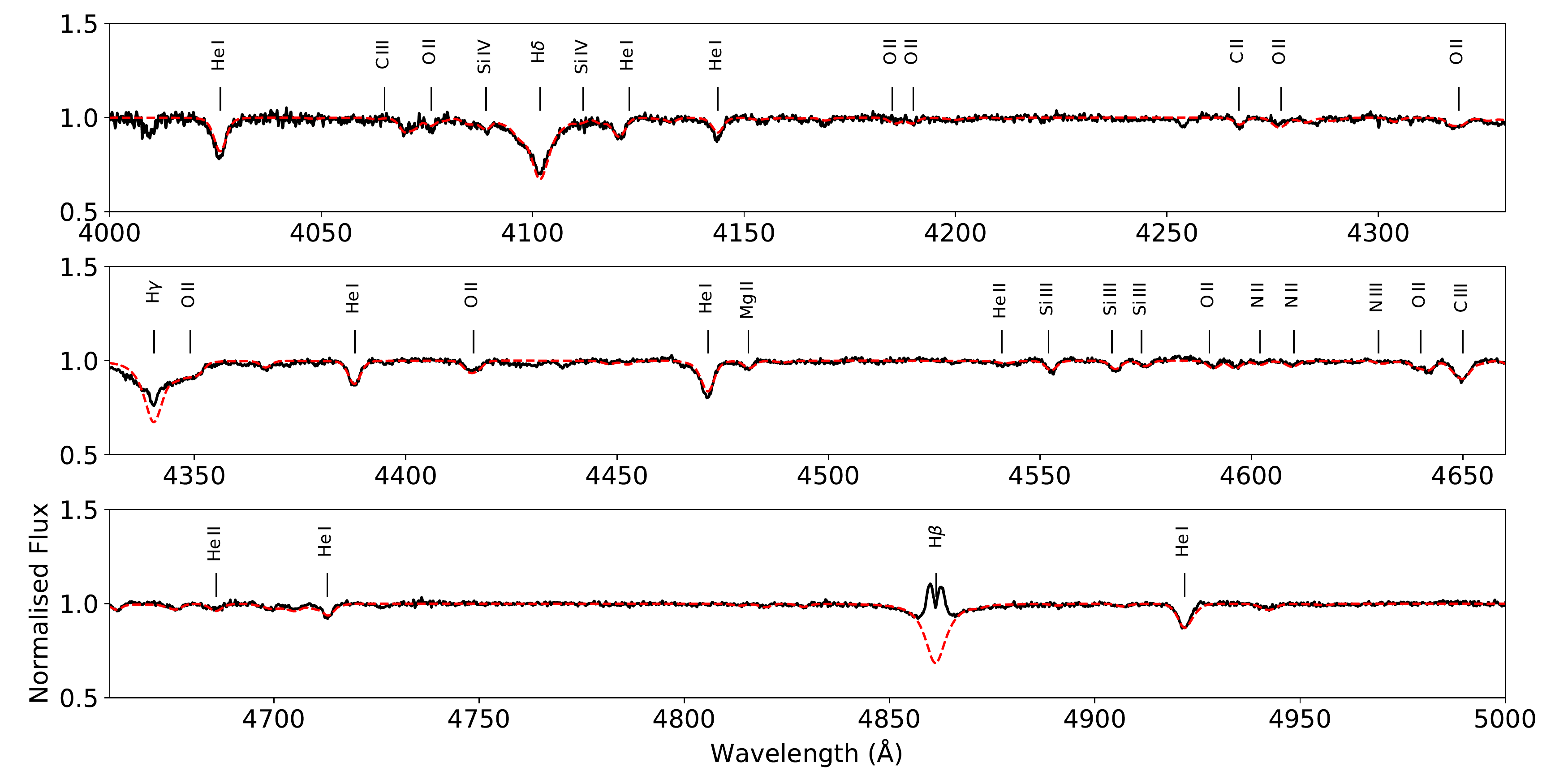}
\end{center}
\caption{Normalized and re-binned spectrum of BSDL\,923 (black line) compared with the best-fitting 
{\sc fastwind} model (red dashed line). The lines used to determine the stellar parameters (shown in 
Tables\,\ref{tab:par} and \ref{tab:abu}) are labeled.
} 
\label{fig:fit}
\end{figure*}

\begin{figure}
    \begin{center}
        \includegraphics[width=0.48\textwidth,angle=0]{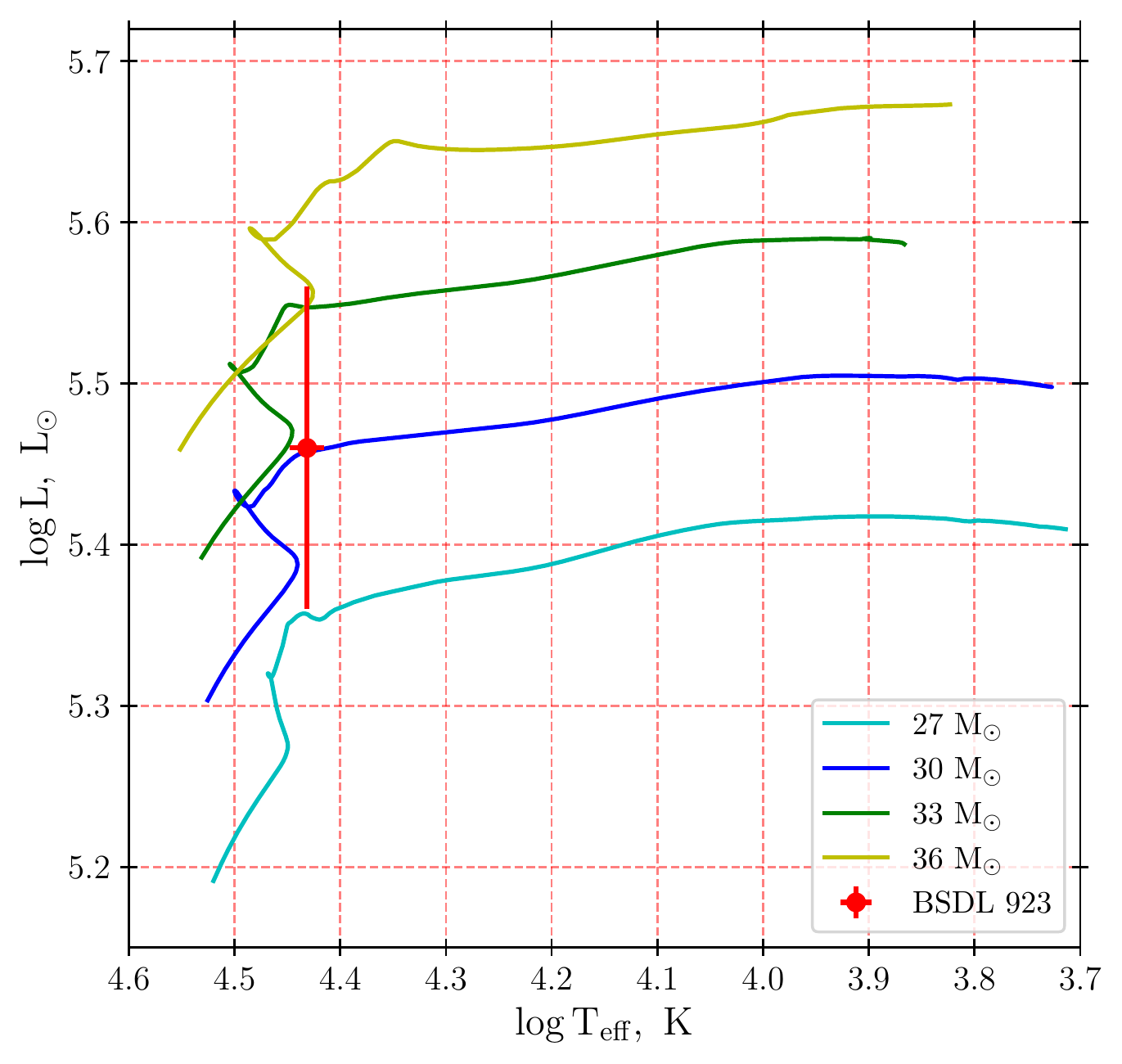}
    \end{center}
    \caption{Position of BSDL\,923 (red symbol) in the Hertzsprung–Russell diagram. 
        The solid lines of different colours show the \mist\ (Choi et al. 2016)   
        evolutionary tracks of a 27--36 M$_\odot$ single, LMC metallicity star.} 
    \label{fig:HR}
\end{figure}

\subsection{Spectral modelling}
\label{sec:par}

Before proceeding with the stellar atmosphere modelling, we combined two of the three LR HRS spectra 
to increase S/N. The spectrum obtained on 2019 October 11 was discarded because it is strongly 
contaminated by emission from the circumstellar medium (see Section\,\ref{sec:cla}). The merged 
spectrum was rebinned reaching S/N of 140 in the spectral range used for modelling (see below).
 
The stellar atmosphere modelling was performed using the stellar atmosphere code {\sc fastwind} 
(Santolaya-Rey, Puls \& Herrero 1997; Puls et al. 2005). We followed the same technique and the 
stellar grid, for LMC metallicity, as described in Castro et al. (2012). Based on the effective 
temperature,  $T_{\rm eff}$, and gravity, $\log g$, derived by the algorithm, a tailored grid, 
exploring a large range of chemical abundances, was built. The stellar parameters (including 
microturbulence $\xi$) and chemical abundances were derived through automatic algorithms searching 
for the set of parameters that best reproduce the main transitions in the range of $\approx 
4000-5000$\,\AA \, (cf. Castro et al. 2012; Gvaramadze et al. 2014, 2019b). 

The projected rotational and macroturbulence velocities, $v\sin i$ (where $i$ is the inclination angle 
between the stellar rotational axis and the line-of-sight) and $v_{\rm mac}$, were measured using the 
He\,{\sc i} $\lambda$4387 line and the {\sc iacob-broad} code (Sim\'on-D\'iaz \& Herrero 2007, 2014). 
We got $v\sin i=100\pm45 \, \kms$ and $v_{\rm mac}=150\pm50 \, \kms$.

The colour excess, $E(B-V)$, and luminosity, $L_*$, of BSDL\,923 were calculated using the photometry 
from Table\,\ref{tab:det} and the synthetic {\sc fastwind} spectral energy distribution, and adopting 
a distance to the LMC of 49.9 kpc (Pietrzy\'nski et al. 2013). We applied the extinction curves 
published by Fitzpatrick \& Massa (2007) until the observed photometry was reproduced. 

The best-fitting model for BSDL\,923 is overlaid on the normalized observed spectrum in 
Fig.\,\ref{fig:fit}, while the stellar parameters and abundances derived in the model are given in 
Tables\,\ref{tab:par} and \ref{tab:abu}.

\begin{table}
    \centering\caption{Parameters of BSDL\,923.}
    \begin{tabular}{lc}
        \hline
        Parameter  & Value \\
        \hline                                         
        $T_{\rm eff}$ (K)             & $27\,000\pm1000$ \\  
        $\log g$                      & $3.22\pm0.10$   \\  
        $\xi \, (\kms)$               & $3\pm1$ \\            
        $v_{\rm macro} \, (\kms)$     & $150\pm50$ \\   
        $v\sin i \, (\kms)$           & $100\pm45$ \\  
        $E(B-V)$ (mag)                & $0.53\pm0.05$ \\  
        $\log (L_*/\lsun)$            & $5.46\pm0.10$ \\
        \hline
    \end{tabular}
    \label{tab:par}
\end{table}
%
\begin{table}
\caption{Metal abundances (by number) in BSDL\,923. 
The present-day chemical composition of the LMC from Hunter et al. (2007) are given for 
reference.}
\label{tab:abu}
\centering
\begin{tabular}{lcc}
\hline $\log(X/{\rm H})+12$ & BSDL\,923 & LMC \\
\hline 
C & $7.69\pm0.15$ & 7.75 \\
N & $7.08\pm0.20$ & 6.90 \\
O & $8.76\pm0.10$ & 8.35 \\
Mg & $7.53\pm0.20$ & 7.05 \\
Si & $7.51\pm0.15$ & 7.20 \\
\hline
\end{tabular}
\end{table}

The colour excess and luminosity can also be derived in an independent way. To estimate $E(B-V)$ one 
can use the $B$ and $V$ photometry from Table\,\ref{tab:det} and the intrinsic colour of a B0.7\,III 
star of $(B-V)_0=-0.25$\,mag (extrapolated from Martins \& Plez 2006), i.e. $E(B-V)=(B-V)-(B-V)_
0=0.55\pm0.05$\,mag, which agrees well with the value given in Table\,\ref{tab:par}. To derive the 
luminosity, we first estimate the absolute visual magnitude of BSDL\,923: $M_V=V-{\rm DM}-R_V E(B-V)$, 
where DM is the distance modulus of the LMC and $R_V$ is the total-to-selective absorption ratio. 
Using DM=18.49\,mag (Pietrzy\'nski et al. 2013) and $R_V=3.1$, one finds 
$M_V=-6.74\pm0.16$\,mag. Taking a bolometric correction of $-2.63\pm0.08$ mag (Lanz \& Hubeny 2007) 
the luminosity of BSDL\,923 is $\log(L_*/\lsun)=5.64\pm0.10$, which agrees within the error margins 
with the value given in Table\,\ref{tab:par}. 
Fig.~\ref{fig:HR} shows the position of BSDL\,923 in the Hertzsprung-Russell diagram 
along with the show the \mist\ (MESA Isochrones and Stellar Tracks; Choi et al. 2016)   
evolutionary tracks of a 27--36 M$_\odot$ single, LMC metallicity star.
This result agrees with mass estimation $M_*=37 ^{+31} _{-17} \, \msun$ presented in Section~6.

\subsection{Spectroscopy of MCSNRJ0513$-$6724}
\label{sec:snr}

\begin{figure}
\begin{center}
\includegraphics[width=8cm,angle=0,clip=]{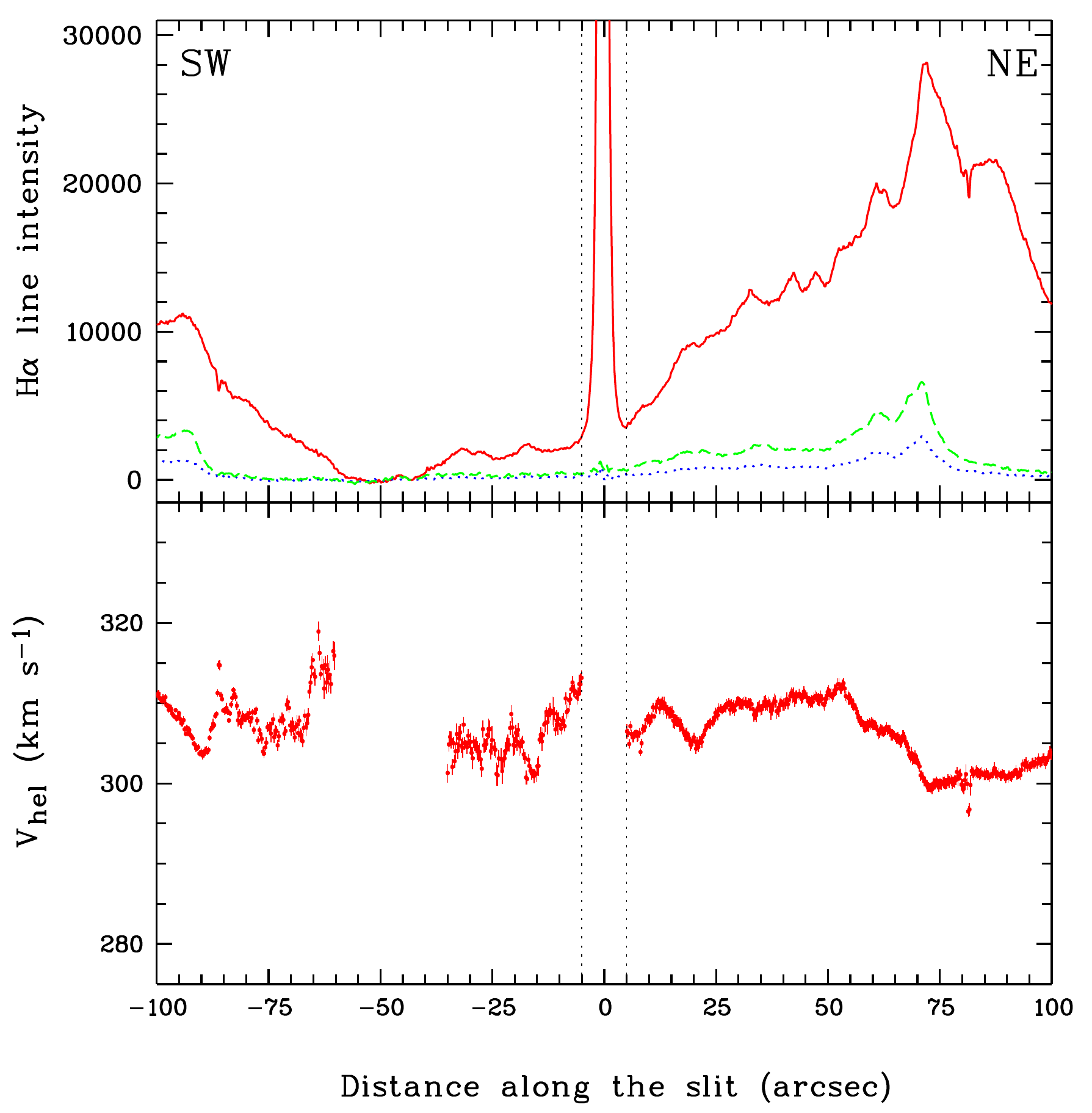}
\end{center}
\caption{Upper panel: H\,$\alpha$, [N\,{\sc ii}] $\lambda$6584 and combined [S\,{\sc ii}] 
$\lambda\lambda$6716, 6731 line intensity profiles along the slit plotted by the (red) 
solid, (green) dashed and (blue) dotted lines, respectively. Bottom panel: H\,$\alpha$ heliocentric 
radial velocity profile along the slit. In both plots MCSNRJ0513$-$6724 occupies the area between 
$\approx-50$ and $+50$ arcsec. Two vertical dashed lines (at $\pm5$ arcsec) mark the area where the 
radial velocity was not measured because of the effect of BSDL\,923. SW--NE direction of the slit is 
shown.} 
\label{fig:ha}
\end{figure}

We found no signs of the SNR shell in the 2D RSS spectrum. Fig.\,\ref{fig:ha} plots the distribution 
of intensity profiles of the H\,$\alpha$, [N\,{\sc ii}] $\lambda$6584 and combined [S\,{\sc ii}] 
$\lambda\lambda$6716, 6731 lines (upper panel), and the H\,$\alpha$ heliocentric radial velocity profile 
(bottom panel) along the slit. In these plots MCSNRJ0513$-$6724 occupies the area between $\approx-50$ 
and $+50$ arcsec. One can see that there is no brightness enhancements or a particularly strong 
[S\,{\sc ii}] emission at the edges of MCSNR\,J0513$-$6724, which is to be expected since this SNR is in 
the adiabatic (non-radiative) phase. 
Also, one can see that the 
heliocentric radial velocity of the H\,$\alpha$ emission towards the SNR does not differ much from
the local systemic velocity  of the LMC of $\approx300 \, \kms$ (see fig.\,3 in Kim et al. 1998), meaning 
that this emission originates in the background \hii region not affected by the SN blast wave.

\section{Discussion}
\label{sec:dis}

The results of spectral modelling could be used to derive the radius and mass of BSDL\,923 through the following 
relations: $R_*=(L_*/4\upi\sigma T_{\rm eff} ^4)^{1/2}$ and $M_* =gR_* ^2/G$, where $\sigma$ and $G$ are, 
respectively, the Stefan-Boltzmann and gravitational constants. Substituting the stellar parameters 
from Table\,\ref{tab:par} into these relations, gives $R_*=25\pm5 \, \rsun$ and $M_*=37 ^{+31} _{-17} \, \msun$. 
With this mass one can calculate the semi-major axis of the NS orbit, $a_{\rm NS}$, and estimate 
the orbital inclination angle $i_{\rm orb}$. The former is given by
\begin{equation}
a_{\rm NS}=\left[{G(M_*+M_{\rm NS})P^2 \over 4\upi ^2}\right]^{1/3} \, ,
\label{eq:axis}
\end{equation}
where $M_{\rm NS}$ is the mass of the NS, while the latter could be determined from the following relation:
\begin{equation}
f(m)={M_{\rm NS} ^3 \sin ^3 i_{\rm orb}\over (M_*+M_{\rm NS})^2} \, ,
\label{eq:mass}
\end{equation}
where the mass function $f(m)$ is given in Table\,\ref{tab:orb}. Adopting the canonical NS mass of $M_{\rm 
NS}=1.4 \, \msun$, one finds from equation\,(\ref{eq:axis}) that $a_{\rm NS}=17\pm3 \, \rsun$. Note that 
this estimate will remain almost the same if $M_{\rm NS}$ is allowed to change within the observed mass 
range for NSs of $\approx1.2-2.0 \, \msun$ (e.g. Martinez et al. 2015; Cromartie et al. 2020). 

\begin{figure}
    \begin{center}
    \includegraphics[width=0.48\textwidth,angle=0]{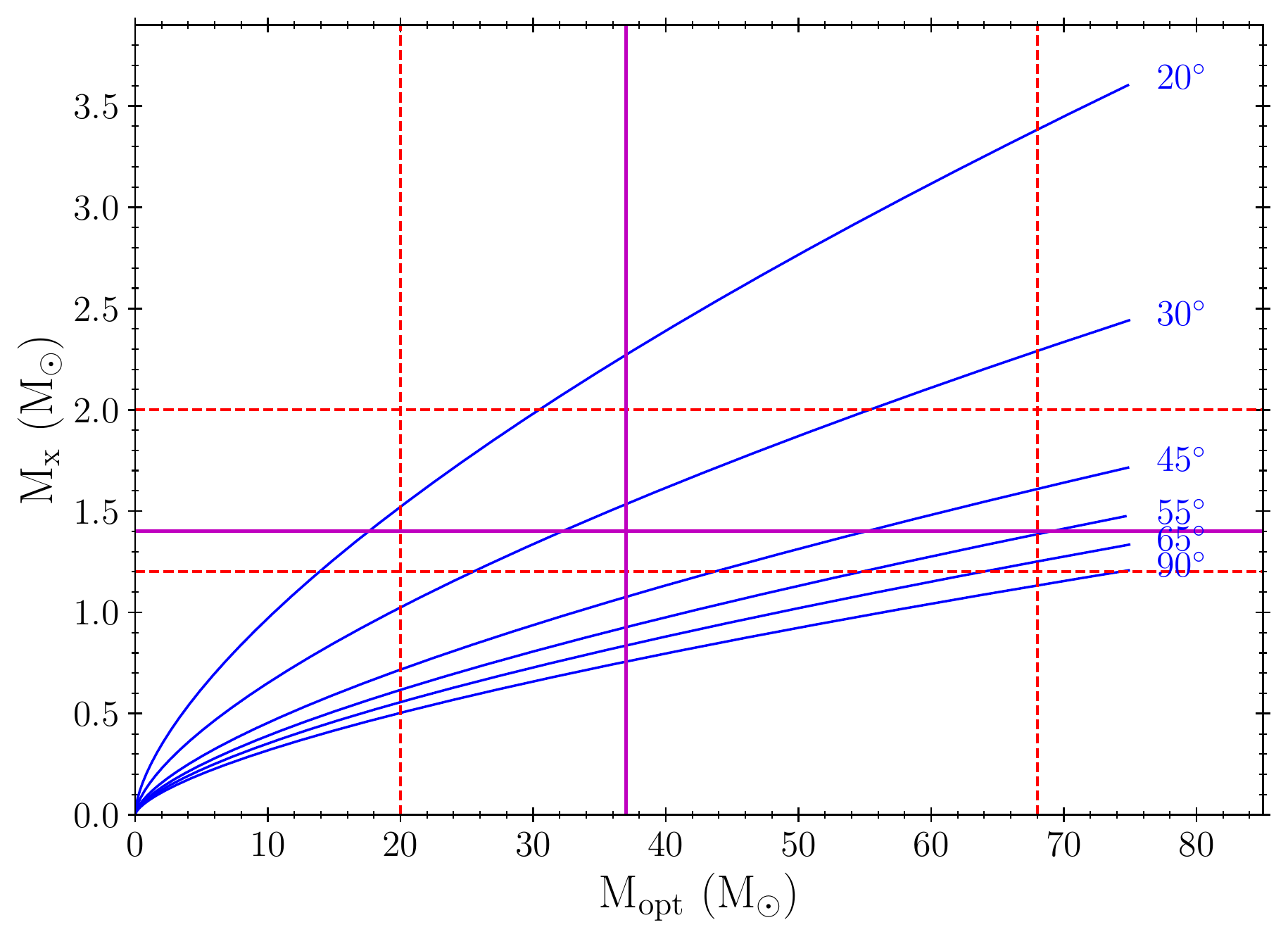}
    \end{center}
    \caption{Mass-mass diagram of XMMU\,J051342.6$-$672412. The (blue) curves show the solution to the 
    mass function for several values of the orbital inclination angle $i_{\rm orb}$. The vertical solid
    line shows the mass of BSDL\,923 of $37 \, \msun$ (with $1\sigma$ uncertainties shown by dashed lines). 
    The horizontal solid line corresponds to a canonical NS mass of $1.4 \, \msun$, while the horizontal
    dashed lines indicate the observed range of NS masses of $1.2-2.0 \, \msun$.}
    \label{fig:angle}
\end{figure}

To constrain the orbital inclination angle $i_{\rm orb}$, we solved equation\,(\ref{eq:mass}) using the 
mass ranges of $20-68 \, \msun$ and $1.2-2.0 \, \msun$ for BSDL\,923 and the NS, respectively. 
Fig.\,\ref{fig:angle} plots solutions to $f(m)$ for several values of $i_{\rm orb}$. For $M_* =37 \, \msun$ 
and $M_{\rm NS}=1.4 \, \msun$, one finds that $i_{\rm orb}\approx 30\degr$, but the uncertainty in the mass 
estimate for BSDL\,923 implies that $i_{\rm orb}$ could range from $\approx 20\degr$ to $55\degr$ (for 
$M_{\rm NS}=1.4 \, \msun$). Assuming that the stellar rotation axis is collinear with the vector of the 
orbital momentum (i.e. $i=i_{\rm orb}$)\footnote{We caution that this may not be the case if the NS received 
a kick at birth.}, one finds that the surface rotational velocity of BSDL\,923 could be as large as 
$\approx300 \, \kms$ (if $i_{\rm orb}=20\degr$), which is not unheard of for OB stars of similar surface 
gravity (see fig.\,1 in Brott et al. 2011).

\begin{figure}
    \begin{center}
    \includegraphics[width=0.48\textwidth,angle=0]{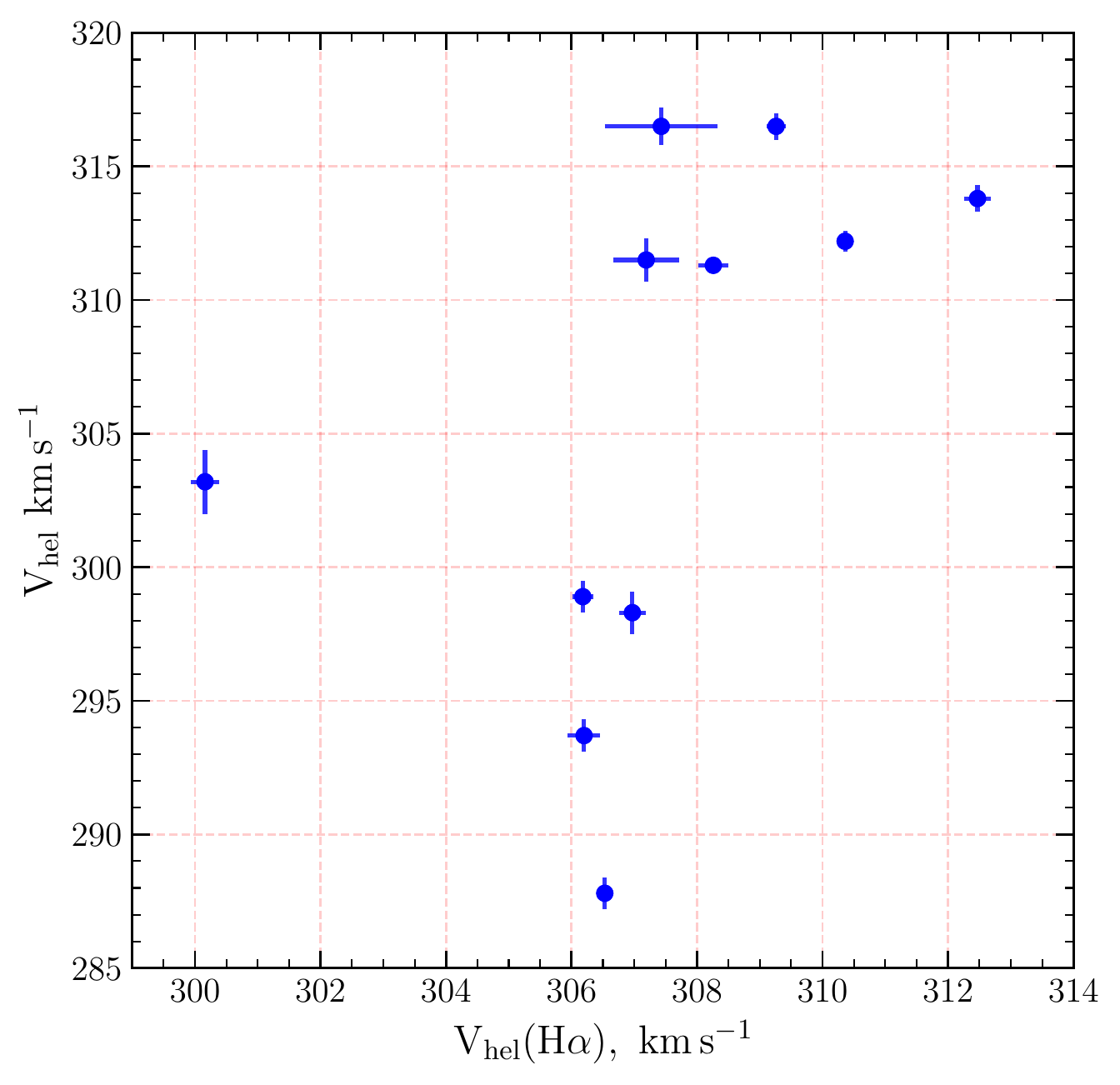}
    \end{center}
    \caption{The H$\alpha$-emitting material velocity versus photospheric velocity 
        (radial velocity of BSDL\,923).
    }
    \label{fig:Vel_v_Vel}
\end{figure}

The non-supergiant nature of BSDL\,923 and the presence of double-peaked emission lines in its spectrum 
is typical of Be stars, implying that XMMU\,J051342.6$-$672412 might be a Be HMXB. If so, then the 
measured EW of the H\,$\alpha$ line of $\approx20-25$ \AA \, (see Table\,\ref{tab:rad}) and the $P_{\rm 
orb}-$EW(H\,$\alpha$) diagram for Be HMXBs (see fig.\,15 in Reig 2011) would imply that $P_{\rm orb}$ of 
XMMU\,J051342.6$-$672412 should be $\approx60-80$\,d, i.e. a factor of $50-60$ longer than what follows 
from our analysis of the radial velocity curve of BSDL\,923. On the other hand, the short spin and orbital 
periods of XMMU\,J051342.6$-$672412 place this HMXB in the leftmost bottom part of the Corbet diagram 
(Corbet 1986; see, e.g., fig.\,2 in Reig 2011) populated by the very rare class of disk-fed supergiant 
HMXBs (currently, only three such objects are known). In these HMXBs the supergiant mass donor star 
overflows its Roche lobe and accretion on the NS occurs through an accretion disc, resulting in high 
(persistent) X-ray luminosity of $L_{\rm X}\sim 10^{38} \, {\rm erg} \, {\rm s}^{-1}$. The X-ray luminosity 
of XMMU\,J051342.6$-$672412 of $\sim10^{34} \, {\rm erg} \, {\rm s}^{-1}$ (Maitra et al. 2019), however, 
is about four orders of magnitudes lower and is more typical of wind-fed HMXBs, in which NSs accrete matter 
directly from the stellar wind. 
However, the Corbet diagram and $\rm P_{spin}-P_{orbit}$ relation are relevant for systems 
near the spin equilibrium which is far from the case for such a young system as BSDL\,923 (see below).

We note that the radial velocity measurements (see Section\,\ref{sec:orb}) show that the 
H\,$\alpha$-emitting material does not participate in the orbital motion of BSDL\,923. This indicates 
that this material does not form an excretion disk around BSDL\,923 (i.e. this star is not a Be star), 
but is rather a rotating circumbinary disk-like structure.
We plotted the H$\alpha$-emitting material velocity versus photospheric velocities of BSDL\,923
(both are presented in from Table~\ref{tab:rad}) in Figure~\ref{fig:Vel_v_Vel}. 
The correlation coefficient between these two quantities is 0.47, 
which corresponds to a weak correlation.

We found that the radius of BSDL\,923 is larger than the semi-major axis of the NS orbit, 
meaning that the NS is enveloped by BSDL\,923. Moreover, since $M_{\rm NS} \ll M_*$, the location of the second 
Lagrangian point L2 is approximately defined by the $R_{\rm L2}\approx a_{\rm NS}\left(1+{\sqrt[{3}]{M_{\rm 
NS}/3M_*}}\right) \approx 1.23a_{\rm NS}$, i.e. also inside BSDL\,923. Such a situation, known as the common 
envelope phase (Paczynski 1976), usually occurs at advanced stages of evolution of the mass donor star in a 
binary system. But it could happen prematurely if the newly formed NS star received a kick of proper orientation 
and amplitude to become embedded in the secondary star (e.g. Leonard, Hills \& Dewey 1994). Alternatively, 
the new-born NS could be engulfed in the secondary star due to inflation of its envelope caused by the energy 
input from the SN blast wave (e.g. Ogata, Hirai \& Hijikawa 2021; cf. Johnston, Soria \& Gibson 2016). The 
energy deposited in the star increases its radius and luminosity, and makes it underdense and cooler than a 
pristine star of the same initial mass and evolutionary phase. The time-scale for thermal adjustment (the 
Kelvin-Helmholtz time-scale) of such a perturbed star, $\tau_{\rm KH}\approx(GM_* ^2/2R_*L_*)\approx 3000$ yr, 
is comparable to the age of the SNR, meaning that BSDL\,923 could still remain bloated. In both cases, such a 
system will ultimately evolve in a Thorne-$\dot{\rm Z}$ytkow object 
(Thorne \& $\dot{\rm Z}$ytkow 1975; Leonard et al. 1994; Hutilukejiang et al. 2018).

Which of the above two situations took place in XMMU\,J051342.6$-$672412 is difficult to say, but in any 
case, it can be expected that the matter lost by BSDL\,923 should concentrate near the orbital plane forming 
a rotating disk-like structure responsible for the origin of two-peaked emission lines in the spectrum of 
this star. Since the common envelope is not rotating as a rigid body (Paczynski 1976), one can expect that at 
least part of the material beyond L2 remains gravitationally bound to the system and that the rotational 
period of this circumbinary material is longer than the binary orbital period. It is therefore tempting to 
assume that the period of $\approx2.23$\,d detected in both the OGLE and TESS light curves of BSDL\,923 is 
associated with the rotational period of the circumbinary material. 

Unlike the excretion disks in Be HMXBs (whose size is defined by the feedback from the 
companion NS), the size of the circumbinary disk-like structure does not depend on the orbital 
period of the binary system, but could be limited by the high thermal pressure of gas in the interior 
of the SNR (cf. Gvaramadze et al. 2021).

Note that the short orbital period of the HMXB and the young age of the SNR (only several thousand years; 
see Section\,\ref{sec:obs}) imply that BSDL\,923 may have peculiar chemical abundances due to pollution 
by SN ejecta. Interestingly, our spectral analysis did reveal enhanced abundances of several
$\alpha$-elements in the atmosphere of this star. Indeed, inspection of Table\,\ref{tab:abu} shows that 
the measured abundances of carbon and nitrogen in BSDL\,923 are in good agreement with the present-day 
abundances of these elements in the LMC, while abundances of the $\alpha$-elements O, Mg and Si are 
$\approx2.6\pm0.6$, $3.0^{+1.8} _{-1.1}$ and $2.0^{+0.9} _{-0.5}$ times enhanced. We interpret this as 
an indication that the surface of the star is contaminated by heavy elements from SN ejecta.

In conclusion, we note that a non-zero value of the orbital inclination angle means that 
the NS must be periodically eclipsed by BSDL\,923. Correspondingly, the X-ray pulsations from the NS 
should be observable only at certain orbital phases before and after the superior conjunction (the 
duration of the eclipse phase depends on the orbital inclination angle and the density profile of the 
atmosphere of BSDL\,923). Unfortunately, the relatively large uncertainty in $P$ and the long time that 
has passed since the X-ray observations (lasting about a third of the orbital period) did not allow us 
to estimate the range of orbital phases covered by {\it XMM-Newton}. 

\section{Summary}
\label{sec:sum}

We have presented the results of high-resolution spectroscopic observations of the mass donor star 
BSDL\,923 in the NS HMXB XMMU\,J051342.6$-$672412 associated with the LMC SNR MCSNR\,J0513$-$6724 
carried out with the Southern African Large Telescope (SALT). We obtained 11 optical \'echelle spectra 
of BSDL\,923, showing a distinct radial velocity variability, which reflects motion of the star in 
an eccentric ($e=0.158\pm0.061$) orbit with a period of $1.280\pm0.006$\,d. Our observations also 
showed that BSDL\,923 is a B0.7\,III star with a double peaked emission spectrum, and that despite 
the orbital motion of BSDL 923 the heliocentric radial velocity of the H\,$\alpha$ emission line 
(originating in the circumstellar medium) remains approximately equal to the systemic velocity of the 
binary system. This indicates that the double peaked emission lines originate in a rotating 
circumbinary disk-like structure.

Using the stellar atmosphere code {\sc fastwind}, we derived some parameters of BSDL\,923, such as 
its effective temperature, $T_{\rm eff}=27\,000\pm100$\,K, surface gravity, 
$\log g=3.22\pm0.10$, projected rotational velocity, $v\sin i \approx100\pm45 \, \kms$, colour excess, 
$E(B-V)=0.53\pm0.05$ mag, and luminosity $\log(L_*\lsun)=5.46\pm0.10$. Using these figures,
we estimated the radius and mass of BSDL\,923 to be, respectively, equal to $R_*=25\pm5 \, \rsun$ and 
$M_*=37 ^{+31} _{-17} \, \msun$, which for the NS mass of $1.4 \, \msun$ gives the semi-major axis of 
the NS orbit of $a_{\rm NS}=17\pm3 \, \rsun$, indicating that the NS is embedded in the atmosphere of 
BSDL\,923. We have speculated that the NS find itself within the atmosphere of BSDL\,923 either because
it was kicked at birth towards the companion star or it became engulfed by the secondary star due to its 
inflation caused by the energy input from the SN blast wave. Another consequence of the encounter 
between the SN blast wave and BSDL\,923 is that the stellar surface should be polluted by SN ejecta. 
Indeed, our spectral modelling has shown that abundances of the $\alpha$-elements O, Mg and Si are 
enhanced by a factor of 2--3 compared to the present-day chemical composition of the LMC.

Also, we have carried out long-slit spectroscopy with SALT of the field containing MCSNR\,J0513$-$6724 in 
order to search for possible optical counterpart to the X-ray shell of the SNR and/or signs of interaction 
between the SN blast wave and the local interstellar medium. We did not find any brightness enhancement or 
a particularly strong [S\,{\sc ii}] emission in place of the X-ray shell which conforms with the young age 
(several thousand years) of MCSNR\,J0513$-$6724, implying that this SNR is in the adiabatic (non-radiative) 
phase.

\section{Acknowledgements}
This work is based on observations obtained with the Southern African Large Telescope (SALT), 
programmes 2019-1-MLT-002 and 2019-1-SCI-009, 2020-1-MLT-003, 
was supported by the National Research Foundation (NRF) of South Africa.
AYK and IYK acknowledge the Ministry of Science and Higher Education of the Russian Federation grant 075-15-2022-262 (13.MNPMU.21.0003).
NC acknowledges funding from the Deutsche Forschungsgemeinschaft (DFG) - CA 2551/1-1. 
This research has made use of the SIMBAD data base 
and the VizieR catalogue access tool, both operated at CDS, Strasbourg, France.

This paper is dedicated to the memory of \mbox{Vasilii Gvaramadze}.

\section{Data availability}

The data underlying this article will be shared on reasonable request to the corresponding author.


\begin{thebibliography}{}
%
\bibitem{} Balona L.~A., 2016, MNRAS, 457, 3724    
\bibitem{} Barnes S. I. et al., 2008, in McLean I. S., Casali M. M., eds, Proc. SPIE Conf. Ser. Vol. 
7014, Ground-based and Airborne Instrumentation for Astronomy II. SPIE, Bellingham, p. 70140K
\bibitem{} Bhattacharya D., van den Heuvel E. P. J., 1991, Phys. Rep., 203, 1
\bibitem{} Bozzetto L. M. et al., 2017, ApJS, 230, 2
\bibitem{} Bramall D. G. et al., 2010, in McLean I. S., Ramsay S. K., Takami H., eds, Proc. SPIE Conf. 
Ser. Vol. 7735, Ground-based and Airborne Instrumentation for Astronomy III. SPIE, Bellingham, p. 77354F
\bibitem{} Bramall D. G. et al., 2012, in McLean I. S., Ramsay S. K., Takami H., eds, Proc. SPIE Conf. 
Ser. Vol. 8446, Ground-based and Airborne Instrumentation for Astronomy IV. SPIE, Bellingham, p. 84460A
\bibitem{} Brott I. et al., 2011, A\&A, 530, A115
\bibitem{} Buckley D. A. H., Swart G. P., Meiring J. G., 2006, in Stepp L. M., ed., Proc. SPIE Conf. Ser. 
Vol. 6267, Ground-based and Airborne Telescopes. SPIE, Bellingham, p. 62670Z
\bibitem{} Burgh E. B., Nordsieck K. H., Kobulnicky H. A., Williams T. B., O'Donoghue D., Smith M. P., 
Percival J. W., 2003, in Iye M., Moorwood A. F. M., eds, Proc. SPIE Conf. Ser. Vol. 4841, Instrument 
Design and Performance for Optical/Infrared Ground-based Telescopes. SPIE, Bellingham, p. 1463
\bibitem{} Castro N. et al., 2012, A\&A, 542, A79
\bibitem{} Choi J. et al., 2016, ApJ, 823, 102
\bibitem{} Christodoulou D. M., Laycock S. G. T., Kazanas D., 2018, MNRAS, 478, 3506
\bibitem{} Corbet R. H. D., 1986, MNRAS, 220, 1047
\bibitem{} Corbet R. H. D. et al., 2016, ApJ, 829, 105
\bibitem{} Crause L. A. et al., 2014, in Ramsay S. K., McLean I. S., Takami H., eds, Proc. SPIE Conf. 
Ser. Vol. 9147, Ground-based and Airborne Instrumentation for Astronomy V. SPIE, Bellingham, p. 91476T
\bibitem{} Crawford S. M. et al., 2010, in Silva D. R., Peck A. B., Soifer B. T., Proc. SPIE Conf. Ser. 
Vol. 7737, Observatory Operations: Strategies, Processes, and Systems III. SPIE, Bellingham, p. 773725
\bibitem{} Cromartie H. T. et al., 2020, Nature Astron., 4, 72
\bibitem{} Dachs J., 1972, A\&A, 18, 271
\bibitem{} Evans C. J. et al., 2015, A\&A, 574, A13
\bibitem{} Fitzpatrick E. L., Massa D., 2007, ApJ, 663, 320
\bibitem{} Fu L., Li X.-D., 2012, ApJ, 757, 171
\bibitem{} Gvaramadze V. V., Miroshnichenko A. S., Castro N., Langer N., Zharikov S. V., 2014, MNRAS, 437, 2761
\bibitem{} Gvaramadze V. V. et al., 2017, Nature Astron., 1, 0116
\bibitem{} Gvaramadze V. V., Kniazev A. Y., Oskinova L. M., 2019a, MNRAS, 485, L6
\bibitem{} Gvaramadze V. V., Kniazev A. Y., Castro N., Grebel E. K., 2019b, AJ, 157, 53
\bibitem{} Gvaramadze V. V., Kniazev A. Y., Gallagher J. S., Oskinova L. M., Chu Y.-H., Gruendl R. A., 
Katkov I. Y., 2021, MNRAS 503, 3856
\bibitem{} Haberl F., Sturm R., Filipov\'ic M. D., Pietsch W., Crawford E. J., 2012, A\&A, 537, L1
\bibitem{} Heinz S. et al., 2013, ApJ, 779, 171
\bibitem{} H\'enault-Brunet V. et al., 2012, MNRAS, 420, L13
\bibitem{} Hills J. G., 1983, ApJ, 267, 322
\bibitem{} Ho W. C. G., Wijngaarden M. J. P., Andersson N., Tauris T. M., Haberl F., 2020, MNRAS, 494, 44 
\bibitem{} Howarth I. D., 1983, MNRAS, 203, 301
\bibitem{} Huang C.H. et al., 2020a, Res. Notes Am. Astron. Soc., 4, 204
\bibitem{} Huang C.H. et al., 2020b, Res. Notes Am. Astron. Soc., 4, 206
\bibitem{} Hunter I. et al., 2007, A\&A, 466, 277
\bibitem{} Hutilukejiang B., Zhu C., Wang Z., L\"u G., 2018, JA\&A, 39, 21
\bibitem{} Israelian G., Rebolo R., Basri G., Casares J., Martin E. L., 1999, Nature, 401, 142
\bibitem{} Johnston H. M., Soria R., Gibson J., 2016, MNRAS, 456, 347
\bibitem{} Kim S., Staveley-Smith L., Dopita M. A., Freeman K. C., Sault R. J., Kesteven M. J., McConnell 
D., 1998, ApJ, 503, 674
\bibitem{} Kniazev A. Y., 2022, AstBu, 77, 334
\bibitem{} Kniazev A. Y., Gvaramadze V. V., Berdnikov L. N., 2016, MNRAS, 459, 3068
\bibitem{} Kniazev A. Y., Usenko I. A., Kovtyukh V. V., Berdnikov L. N., 2019, Astrophys. Bull., 74, 208
\bibitem{} Kniazev A. Y., Malkov O. Y., Katkov I. Y., Berdnikov L. N., 2020, Res. Astron. Astrophys., 20, 119
\bibitem{} Kniazev A. Y., 2020, Ap\&SS, 365, 169
\bibitem{} Kobulnicky H. A., Nordsieck K. H., Burgh E. B., Smith M. P., Percival J. W., Williams T. B., 
O'Donoghue D., 2003, in Iye M., Moorwood A. F. M., eds, Proc. SPIE Conf. Ser. Vol. 4841, Instrument Design 
and Performance for Optical/Infrared Ground-based Telescopes. SPIE, Bellingham, p. 1634
\bibitem{} Lanz T., Hubeny I., 2007, ApJS, 169, 83
\bibitem{} Leonard P. J. T., Hills J. G., Dewey R. J., 1994, ApJ, 423, L19
\bibitem{} Linares M. et al., 2010, ApJ, 719, L84
\bibitem{} Lomb N. R., 1976, Ap\&SS, 39, 447
\bibitem{} Lozinskaya T. A., 1992, Supernovae and Stellar Wind in the Interstellar Medium. Am. Inst. Phys., New York
\bibitem{} Maitra C. et al., 2019, MNRAS, 490, 5494
\bibitem{} Maitra C., Haberl F., Maggi P., Kavanagh P., Vasilopoulos G., Sasaki M., Filipovic M. D., Udalski A.,
2021, MNRAS, 504, 326
\bibitem{} Martinez J.G. et al., 2015, ApJ, 812, 143
\bibitem{} Martins F., Plez B., 2006, A\&A, 457, 637
\bibitem{} Maryeva O. V., Gvaramadze V. V., Kniazev A. Y., Berdnikov L. N., 2020, MNRAS, 498, 5093
\bibitem{} McBride V. A., Coe M. J., Negueruela I., Schurch M. P. E., McGowan K. E., 2008, MNRAS, 388, 1198
\bibitem{} Meixner M. et al., 2006, AJ, 132, 2268
\bibitem{} O'Donoghue D. et al., 2006, MNRAS, 372, 151
\bibitem{} Ogata M., Hirai R., Hijikawa K., 2021, MNRAS, 505, 2485
\bibitem{} Paczynski B., 1976, in Eggleton P., Mitton S., Whelan J., eds, Proc. IAU Symp. 73, Structure 
and Evolution of Close Binary Systems. Kluwer, Dordrecht, p. 75
\bibitem{} Pennock C.M. et al., 2021, MNRAS, 506, 3540
\bibitem{} Pietrzy\'nski G. et al., 2013, Nature, 495, 76
\bibitem{} Puls J., Urbaneja M. A., Venero R., Repolust T., Springmann U., Jokuthy A., Mokiem M. R., 2005, 
A\&A, 435, 669
\bibitem{} Reig P., 2011, Ap\&SS, 332, 1
\bibitem{} Ricker G. R. et al., 2014, in Oschmann J. M. Jr, Clampin M., Fazio G. G., MacEwen H. A., eds, 
Proc. SPIE Conf. Ser. Vol. 9143, Space Telescopes and Instrumentation 2014: Optical, Infrared, and Millimeter
Wave. SPIE, Bellingham, p. 15
\bibitem{} Santolaya-Rey A. E., Puls J., Herrero A., 1997, A\&A, 323, 488
\bibitem{} Scargle J. D., 1982, ApJ, 263, 835
\bibitem{} Seitenzahl I. R., Vogt F. P. A., Terry J. P., Ghavamian P., Dopita M. A., Ruiter A. J., 
Sukhbold T., 2018, ApJ, 853, L32
\bibitem{} Seward F. D., Charles P. A., Foster D. L.,Dickel J. R., Romero P. S., Edwards Z. I., Perry M., 
Williams R. M., 2012, ApJ, 759, 123
\bibitem{} Shakura N., Postnov K., Kochetkova A., Hjalmarsdotter L., 2012, MNRAS, 420, 216
\bibitem{} Sim\'on-D\'iaz S., Herrero A., 2007, A\&A, 468, 1063
\bibitem{} Sim\'on-D\'iaz S., Herrero A., 2014, A\&A, 562, A135
\bibitem{} Tauris T. M., van den Heuvel E. P. J., 2006, in LewinW. H. G., van der Klis
M., eds, Formation and Evolution of Compact Stellar X-ray Sources,
Compact stellar X-ray sources. Cambridge Univ. Press, Cambridge, p. 623
\bibitem{} Tauris T. M., Fender R. P., van den Heuvel E. P. J., Johnston H. M., Wu K., 1999, MNRAS, 310, 1165
\bibitem{} Thorne K. S., $\dot{\rm Z}$ytkow A. N., 1975, ApJ, 199, L19
\bibitem{} Townsend L. J., Coe M. J., Corbet R. H. D., Hill A. B., 2011, MNRAS, 416, 1556
\bibitem{} van den Heuvel, E. P. J. 2019, in Oskinova L., Bozzo E., Gies D., Holz D., eds, Proc. IAU Symp. 346, 
High-mass X-ray Binaries: Illuminating the Passage from Massive Binaries to Merging Compact Objects.
Cambridge Univ. Press, Cambridge, p. 1
\bibitem{} van Kerkwijk M. H., van Oijen J. G. J., van den Heuvel E. P. J., 1989, A\&A, 209, 173
\bibitem{} van Kerkwijk M. H., van Paradijs J., Zuiderwijk E. J., Hammerschlag-Hensberge G., Kaper L., 
Sterken C., 1995, A\&A, 303, 483
\bibitem{} van Soelen B., Komin N., Kniazev A., V\"ais\"anen P., 2019, MNRAS, 484, 4347
\bibitem{} Wang W., Tong H., 2020, MNRAS, 492, 762 
\bibitem{} Zaritsky D., Harris J., Thompson I. B., Grebel E. K., 2004, AJ, 128, 1606
%
\end{thebibliography}
\end{document}